\documentstyle[12pt,a4]{article}
\makeatletter
\@addtoreset{equation}{section}
\makeatother

\begin{document}
%\maketitle
 
%\begin{titlepage} 
 \begin{center}
{\Large\bf Thermodynamical Bethe Ansatz analysis 
in an SU(2)$\times$U(1) symmetric $\sigma$-model}
\end{center}
\vskip 2.5 true cm
\centerline{
{\large J\'anos Balog}\footnote{on leave from the Research 
Institute for Particle and Nuclear Physics, Budapest, Hungary} 
and {\large P\'eter Forg\'acs}
}
\vskip1ex
\centerline{Laboratoire de Math\'ematiques et Physique Th\'eorique}
\centerline{Universit\'e de Tours}
\centerline{Parc de Grandmont, 37200 Tours, France}
\vskip 3.20 true cm
\centerline{\bf Abstract}
\vskip 1.0ex
Four different types of free energies
are computed by both thermodynamical Bethe Ansatz (TBA) techniques
and by weak coupling perturbation theory
in an integrable one-parameter deformation of the O(4) principal
chiral $\sigma$-model (with SU(2)$\times$U(1) symmetry). 
The model exhibits both \lq fermionic' and \lq bosonic'
type free energies and in all cases the perturbative and the 
TBA results are in perfect agreement,
strongly supporting the correctness of the proposed $S$ matrix.
The mass gap is also computed in terms of 
the $\Lambda$ parameters of the modified minimal substraction
scheme and a lattice regularized version of the model.
\vfill
\eject

%\end{titlepage}
 
%
%
\newcommand{\eff}{\lambda_{\scriptscriptstyle {\rm eff}}}
\newcommand{\bee}{\begin{equation}}
\newcommand{\ee}{\end{equation}}
\newcommand{\ba}{\begin{array}}
\newcommand{\ea}{\end{array}}
\newcommand{\bea}{\begin{eqnarray}}
\newcommand{\eea}{\end{eqnarray}}
\newcommand{\e}{\epsilon_+(i\kappa)}
\newcommand{\eps}{\epsilon}
\newcommand{\pa}{\partial}
\newcommand{\lb}{\lbrack}
\newcommand{\Se}{S_{\rm eff}}
\newcommand{\rb}{\rbrack}
\newcommand{\de}{\delta}
\newcommand{\th}{\theta}
\newcommand{\ka}{\kappa}
\newcommand{\al}{\alpha}
\newcommand{\si}{\sigma}
\newcommand{\vp}{\varphi}
\newcommand{\g}{\gamma}
\newcommand{\om}{\omega}
\newcommand{\pr}{\prime}
\newcommand{\gbb}{\bar{g}}
\newcommand{\gb}{\overline g}
\newcommand{\nb}{\overline N}
\newcommand{\MSb}{{\overline {\rm MS}}}
\newcommand{\lnh}{\ln(h^2/\Lambda^2)}
\newcommand{\df}{\delta f(h)}
\newcommand{\h}{{1\over2}}
\newcommand{\R}{m/\Lambda}
\newcommand{\abschnitt}[1]{\par \noindent {\large {\bf {#1}}} \par}
\newcommand{\subabschnitt}[1]{\par \noindent
                                          {\normalsize {\it {#1}}} \par}
%-----------------------------------------------------------------------
% The definition below makes spaces e.g \skipp{3} makes 3 spaces
\newcommand{\skipp}[1]{\mbox{\hspace{#1 ex}}}
 
%
%
% various slashed symbols
%
%
%\newcommand\slash#1{\rlap{$#1$}/} % slashes a character
\newcommand\dsl{\,\raise.15ex\hbox{/}\mkern-13.5mu D}
    % this one can be subscripted
\newcommand\delsl{\raise.15ex\hbox{/}\kern-.57em\partial}
\newcommand\Ksl{\hbox{/\kern-.6000em\rm K}}
\newcommand\Asl{\hbox{/\kern-.6500em \rm A}}
\newcommand\Dsl{\hbox{/\kern-.6000em\rm D}} %roman D
\newcommand\Qsl{\hbox{/\kern-.6000em\rm Q}}
\newcommand\gradsl{\hbox{/\kern-.6500em$\nabla$}}
%-----------------------------------------------------------------------

\pagestyle{plain}
\setcounter{page}{1}
\setcounter{section}{1}
\abschnitt{1. Introduction}
 
We shall study the following one-parameter family 
of 1+1 dimensional $\sigma$-models described by the Lagrangian
\bee
{\cal L}_0=-{1\over2\lambda}\Big(L^1_\mu L^1_\mu+L^2_\mu L^2_\mu+
    (1+g)L^3_\mu L^3_\mu\Big)\,,
\label{Lag0}
\ee
where
\bee
{1\over2}L^a_\mu\sigma^a=G^{-1}\partial_\mu G\,,\qquad G\in{\rm SU(2)}\,,
\ee
 $g$ is a (real)
parameter and $\sigma^a$ stand for the standard Pauli matrices.
The Lagrangian (\ref{Lag0}) can be interpreted as a deformation of
the SU(2)$\times$SU(2) (or O(4)) symmetric nonlinear $\sigma$-model (NLS)
by the parameter $g$. It possesses an SU(2)$_{\rm L}\times$U(1)$_{\rm R}$ 
(global) symmetry.
We shall assume in the following
that the deformation parameter, $g$, satisfies
\bee
-1\leq g\leq0\,,
\label{ineq1}
\ee
because with $g$ being in this range the model, (\ref{Lag0}), is
asymptotically free (AF) (see Appendix D).
At the classical level the deformed model, (\ref{Lag0}), interpolates
between the SU(2)$\times$SU(2) ($g=0$) and O(3) ($g=-1$) NLS models.
It has already appeared in the Bethe Ansatz approach
of Polyakov and Wiegmann to solve the SU(2)$\times$SU(2) and
O(3) NLS \cite{PW,Wiegmann}. The deformed model (\ref{Lag0})
is known to possess a Lax-pair
and is believed to be integrable both at the classical
and at the quantum level \cite{Cherednik}.
Its $S$ matrix has been proposed long time ago \cite{PW,Wiegmann} and
at present a considerable amount of evidence corroborates the belief
that the deformed SU(2)$\times$U(1) symmetric
 $\sigma$-model is quantum integrable indeed
\cite{Babu, Kirillov, Fateev}. 
Its spectrum is thought to contain two massive doublets
whose scattering is described by the tensor product of
an SU(2)$\times$U(1) symmetric solution of the bootstrap $S$-matrix equations:
\bee
S(\theta)=S^{(\infty)}(\theta)\otimes S^{(p)}(\theta)\,,
\label{Smatrix}
\ee 
where $S^{(p)}(\theta)$ is the Sine-Gordon (SG) $S$-matrix depending on
the parameter $p$. In Eq.\ (\ref{Smatrix}) the limit $p\to\infty$ 
corresponds to an SU(2)$\times$SU(2) symmetric $S$-matrix. 

In this paper we carry out a rather thorough consistency check on the 
proposed $S$-matrix of the one parametric deformation
of the O(4) $\sigma$-model, (\ref{Lag0}).
The basic idea is to investigate the system at large
particle density when both perturbation theory 
in the coupling $\lambda$ and a calculation
based on the $S$--matrix of the massive physical particles
apply. (For a review, see Ref. \cite{hollo}.) To achieve this
one introduces a chemical potential, $h$, coupled to a Noether charge
of the corresponding symmetry,
and considers the ground state of the modified Hamiltonian
\bee \label{hamil}
 H=H_{0} - hQ\,.
\ee
The Legendre transform of the ground state energy density
or (zero temperature) free energy of the system   
must be the form:
\bee \label{free0}
{\cal F}\equiv\df \equiv f(h)-f(0)=-h^2 F_0(h/m,Q)\,,
\ee
where $m$ is the mass of the particles. (Eq.\ (\ref{free0}) follows
simply from dimensional analysis.)
The Thermodynamical Bethe Ansatz (TBA) method 
based on the $S$--matrix leads to a set of integral equations,
which can be solved, for large chemical potential $h$, 
in the form of an asymptotic series expansion 
in $h/m$ using a modified Wiener-Hopf technique \cite{Jap}.

On the other hand in an asymptotically free theory the result of
a perturbative computation of $F_0$
(which is of course renormalization group invariant)
is an asymptotic series in the running coupling,
$\bar\lambda(h/\Lambda)\propto[\ln(h/\Lambda)]^{-1}$,  
where the $\Lambda$ parameter (of dimension mass) 
is the usual renormalization group invariant combination
of the ultraviolet cutoff and the bare coupling, $\lambda$. 
 
Since ${\cal F}$ is a physical quantity, the results obtained by TBA method 
and by perturbation theory should agree.
Thus comparison of the two results
provides a rather stringent consistency check on the
$S$-matrix (and also on the self-consistency of the hypothesis used in the
course of the calculation), moreover 
one can also extract the exact $m/\Lambda$ ratio, a rather important
non-perturbative parameter of the theory which
can then be also measured e.g.\ by lattice simulations.
This comparison, following the
pioneering work in Refs.\ \cite{history, FoNiWe},
has since been applied to many bosonic, fermionic
and supersymmetric integrable models.

In Ref.\ \cite{Fateev} a two-parameter family of integrable
models has been studied by similar techniques. 
The models were mainly treated as perturbed conformal field
theories allowing for a proof of integrability at the quantum
level.
The SU(2)$\times$U(1) symmetric $S$-matrix (\ref{Smatrix}) is  
a limiting case, ($p_1\rightarrow\infty$, $p_2=p$), of the 
$S$-matrix of Ref.\ \cite{Fateev}. 
%Similarly, the kernels
%of the TBA integral equation are also related
%The integral kernels that we used in our TBA integral equation
%were simply the corresponding limits of the
%integral kernels of Ref. \cite{Fateev}.
In the present paper, however, we put the emphasis on the $\sigma$-model
representation suitable for traditional weak-coupling perturbation
theory. Although the model (\ref{Lag0}) corresponds to
a limiting case of the two-parameter family of Ref.~\cite{Fateev},
one cannot read off directly
the asymptotic expansions for the ground state energies 
from Ref.\ \cite{Fateev}. 
This is due to the fact that both in perturbation theory and
using the TBA only the asymptotic series are computable.
The coefficients of these series diverge in the $p_1\to\infty$ limit,
thus one cannot directly recover the coefficients of the limiting case 
from those valid for finite values of the parameters.
For this reason we had to perform the TBA analysis directly in
the $p\to\infty$ limit.

%because the $p_1\rightarrow\infty$
%and the $h\rightarrow\infty$ limits do not commute. 
%Thus we had to
%solve the TBA integral equation with the limit kernels first and
%then expanded the solution asymptotically in $h/m$.
%A completely analogous situation is known to
%occur when one takes
%the $O(4)$ symmetric limit, $p\rightarrow\infty$, of the deformed model 
%(\ref{Lag0}). Here again, the $p\rightarrow\infty$
%and the $h\rightarrow\infty$ limits do not commute. Indeed, the
%first term in the asymptotic expansion of the
%ground state energy of the $O(4)$ symmetric model contains
%a logarithmic factor, which is absent from 
%that of the deformed model for any finite $p$
%and thus cannot be reproduced  in the $p\rightarrow\infty$ limit.

As there are two conserved Noether charges in the model (\ref{Lag0}),
$Q_{\rm L}$, $Q_{\rm R}$,
there are three types of free energies corresponding to ground states
either coupled to $Q_{\rm L}$ or to $Q_{\rm R}$ or to both of them.
In fact when the ground state is coupled to 
$Q_{\rm L}$,  one has to distinguish between two different cases
namely $p>1$ and $p<1$ as then bound states 
(breathers) appear in the spectrum.
We have calculated all the (four) different types of (zero temperature) 
free energies both by the TBA method and by perturbation theory 
in the modified minimal substraction scheme ($\overline{{\rm MS}}$) and
extracted the corresponding (four) $m/\Lambda$ ratios. 
Three of the kernels of the integral equation
determining the ground state energies are simply the
$p_1\to\infty$ limits of the corresponding ones in
\cite{Fateev} whereas  in the $p<1$ case with bound states
we have computed the corresponding kernel directly using
the bootstrap fusion method.
We have found 
complete consistency for all four cases, in particular all 
of the four $m/\Lambda$ ratios agree, the result being:
\bee
{m\over\Lambda_{{\overline{{\rm MS}}}}}=2^{3-{p\over2}}\,
e^{{p\over2}-1}\,
{\Gamma\big(1+{p\over2}\big)\over\pi p}\,,
\label{MperLambda}
\ee
where $m$ is the mass of the SU(2) doublets.
The $p\to0$ limit (corresponding to the running
deformation parameter $\bar g(h)\to-1$ in (\ref{Lag0}) is expected to give the
O(3) NLS model (and a decoupled free field) \cite{Wiegmann}. Our results
are in perfect agreement with his expectation, 
as the previously
calculated $m_1/\Lambda_{{\overline{{\rm MS}}}}=8/e$ 
ratio in the O(3) NLS model \cite{history}
(where now $m_1$ denotes the
mass of the triplet in the O(3) model) is reproduced in the $p\to0$ limit.
This can be immediately seen from Eq.\ (\ref{MperLambda}) using the fact that
in the $p\to0$ limit $m_1\to \pi pm$.   
 
We have also proposed a lattice action for the deformed O(4) NLS model
and calculated the ratio of the $\Lambda$-parameters corresponding to the
two regularization schemes:
\bee
{\Lambda_{\overline{\rm MS}}\over\Lambda_L}=\sqrt{32}
e^{{5\pi\over8}}\,.
\label{Lambdaratio0}
\ee

The plan of the paper is as follows: in Section 2 the calculation of the
free energies by the TBA method is outlined, in Section 3 the perturbative
computation is presented in some detail and the $m/\Lambda$ ratios
are also given there. Section 4 is devoted to the perturbative calculation of
the free energy using lattice regularization leading to the
${\Lambda_{\overline{\rm MS}}/\Lambda_L}$ ratio. 
To make the paper rather self contained we included four
Appendices with some details of the computations.
Appendix A contains some useful
formulae to obtain the free energy by solving the pertinent integral equation.
In Appendix B the scattering phase shift of the lowest mass bound
state is calculated.
Appendix C contains some details of the perturbative calculations, while 
Appendix D is devoted to renormalization group considerations.

\setcounter{section}{2} \setcounter{equation}{0}
\abschnitt{2. Calculation of the free energies from the integral equation}

As the model ({\ref{Lag0}) admits $G=$
\hbox{SU(2)$_{\rm L}\times$U(1)$_{\rm R}$} as its (global) 
symmetry group, one can introduce two chemical potentials coupled to the two
Noether charges, $Q_{\rm L}$, $Q_{\rm R}$, corresponding to the two U(1)
(L and R) subgroups of $G$. The modified Hamiltonian then reads  
 \bee \label{hamil2}
 H=H_{0} - h_{\rm L}Q_{\rm L}-h_{\rm R}Q_{\rm R}\,.
\ee
For definiteness we chose \hbox{$h_{\rm L}, h_{\rm R}\geq0$} and normalize the
charges of the four particles as 
\bee\label{charges}
(1,1)\,,\quad(-1,1)\,,\quad(1,-1)\quad(-1,-1)\,.
\ee
We distinguish between the three different types
of finite density ground states depending on $h_{\rm L}, h_{\rm R}$:
\begin{itemize}
\item[1.] $h_{\rm L}\,, h_{\rm R}>0\,,\!\qquad\qquad$  (DIAG)
\item[2.] $h_{\rm L}=0$\,, $h_{\rm R}=h>0\,,$ (RIGHT)
\item[3.] $h_{\rm L}=h>0$\,, $h_{\rm R}=0\,.$ (LEFT)
\end{itemize}
In the following we shall analyse these three cases.
\subabschnitt{2.1 The diagonal current}
We recall that our strategy will be to calculate
the ground state energy density of the system at large particle density 
(and at zero temperature), that is
we consider the system  
with $h_{\rm L}/m$, $h_{\rm R}/m\gg1$ ($m$ is the mass of the particles).
Then it is clear that particles of charge (1,1) condense into the vacuum.
It is less clear what other kind of particles (necessarily with a smaller
charge/mass ratio) will appear in the vacuum state. We shall assume
that the vacuum consist of {\sl only} particles of charge (1,1).
This seemingly radical assumption
has apparently worked in all analogous examples studied so far,
and it greatly simplifies the solution of TBA equations.

The scattering phase of the (1,1) type particle, $\delta(\theta)$, 
can be read off from Eq. ({\ref{Stensor}) in Appendix B:
\bee\label{diagphase}
\delta(\theta)=\delta_\infty(\theta)+\delta_p(\theta)\,.
\ee
When the ground state is assumed to contain {\sl only} particles of charge (1,1)
the calculation of the free energy
using the TBA method reduces to the solution
of a single integral equation (\ref{inteq}) where the kernel $K$ is given by
the logarithmic derivative of the relevant $S$-matrix element,
\bee
K(\theta)={1\over2\pi i}{d\over d\theta} \ln S(\theta)\,.
\ee
In the present case $S(\theta)=\exp{\{i\delta(\theta)\}}$ 
where scattering phase, $\delta(\theta)$, is given by Eq.\ (\ref{diagphase}). 
We note that the effective chemical 
potential, $h$, in Eq.\ (\ref{inteq}) is just 
$h=h_{\rm L}+ h_{\rm R}$. From Eqs.\ (\ref{diagphase}, \ref{pkernel}) 
the Fourier transform of the relevant
kernel is easily found to be:
\bee \label{kernel}
 1-\hat{K}_{\rm D}(\om)={ \tanh{\pi\om\over2}\over
 2\sinh{\pi p\om\over2} }\, e^{p\pi\vert\om\mid\over2}\,,
\ee
with $p>0$.
The result of the splitting of 
$ 1-\hat{K}_{\rm D}(\om)= \lb G_+(\om) G_-(\om) \rb^{-1}$ where
$G_+(\om)$ resp.\ $G_-(\om)$ are analytic functions in the upper resp.\
lower half plane can be written as
\bee \label{gamma}
 G_+(\om)= \sqrt{2p}\,{\Gamma^2(1-{i\om\over2})\over
\Gamma(1-i\om)\Gamma(1-{ip\om\over2}) }\,
e^{ {i p\om\over2}\left(1-\ln(-{ip\om\over2})\right) }
 e^{ -i\om\ln2 } \,,
\ee
and $G_-(\om)= G_+(-\om)$.

For the present (fermionic type) kernel (\ref{kernel}) the constants in
Eqs.~(\ref{fren1}-\ref{tildeA}) of Appendix~A are given as
\bea
&\tilde a=-{p\over2}\,,\qquad
&\tilde b={p\over2}\left(\ln{p\over2}-1-\Gamma'(1)\right)+\ln2\,,\\
&\tilde k=\sqrt{2p}\,,\qquad
&G_+(i)={\pi\over2}{\sqrt{2p}\over\Gamma(1+{p\over2})}
e^{ -{p\over2}(1-\ln{p\over2}) }\nonumber\,.
\label{const}
\eea
Then the expression for free energy (\ref{fren1}) finally reads:
\bee \label{ferfreen}
\delta{f}_{\rm D}(h)=-{ h^2 \over\pi}\,p
\left\lb 1-{p\over2t}-
{p^2\over4}{\ln t\over t^2}+
{\tilde A \over t^2}+O({\ln t \over t^3}) \right\rb\,,
\ee
where $t=\ln(h/m)$ and the constant $\tilde A$,
which enters in our final result for the $m/\Lambda$ ratio is given as
\bee \label{atild}
\tilde A={p\over2}\left[\ln\Gamma(1+{p\over2})+{p\over2}({3\over2}-\ln2)
-1+3\ln2 -\ln\pi\right]\,.
\ee
We remark here that from the classical term in Eq.\ (\ref{ferfreen}),
$-h^2p/\pi$, one can immediately read off the level of an underlying
ultraviolet (UV) current algebra, $k$, as $k=p$ \cite{fendley}.
%\newpage

\subabschnitt{2.2 The right current}
In analogy to the assumption made in the diagonal case we now expect
the ground state to consist of a mixture of
particles of the same charge with respect to U(1)$_{\rm R}$,
i.e.\ those of charge (1,1) and (-1,1) with equal
densities. In Ref.\ \cite{Fateev} the result of the
diagonalisation of the pertinent coupled TBA system has been given and
the relevant kernel, $\hat K_{\rm R}(\omega)$ turns out to be
the $p_1\to\infty$
limit in Eq.\ (57) of Ref.\ \cite{Fateev}. From this one finds 
(somewhat surprisingly)
$\hat K_{\rm R}(\omega)=\hat K_{\rm D}(\omega)$,
implying that $\delta f_{\rm R}(h)=\delta f_{\rm D}(h)$,
where we have denoted the chemical potential for the right current as
$h_{\rm R}=h$.
So the free
energies are precisely the same as functions of the effective chemical
potentials in these two apparently rather different cases.
We do not have an explanation of this fact,
we note, however, that as long as $p_1<\infty$  
$\hat K_{\rm R}(\omega)\ne\hat K_{\rm D}(\omega)$.

\subabschnitt{2.3 The left current}
Analogously to the RIGHT case
the ground state is expected to consist of an
equal density mixture of
particles of the same charge, this time with respect to U(1)$_{\rm L}$,
i.e.\ those of charge (1,1) and (1,-1).
Using again the result of Ref.\ \cite{Fateev}
the Fourier transform of the relevant (bosonic type) kernel
is found to be:
\bee \label{leftkernel}
 1-\hat{K}_{\rm L}(\om)={ \tanh{\pi\mid\om\mid\over2}\over
2\cosh{\pi x\om\over2} }\,
e^{x\pi\mid\om\mid\over2}\,,
\ee
where $x=p-1$ and $x$ is assumed to be {\sl positive}.
The functions appearing in the decomposition of
$ 1-\hat{K}_{\rm L}(\om)$ are now given as
\bee \label{leftgamma}
 G_+(\om)= {2\sqrt{\pi}\over\sqrt{-i\om}}
 { \Gamma(1-{i\om\over2})\over
\Gamma(\h-{i\om x\over2})\Gamma(\h -{i\om\over2}) }\,
e^{ {i x\om\over2}\left(1-\ln(-{ix\om\over2})\right)\,, } 
\ee 
and $G_-(\omega)=G_+(-\omega)$.
For the kernel (\ref{leftkernel}) the constants in Eq.\ (\ref{leftfreen})
are not difficult to calculate and we find:
\bea
&a=-{x\over2}\,,\qquad
&b=-\h \left[x\ln x +x(\ln2-1-\Gamma'(1))+2\ln2\right]\,,\nonumber\\
&k={2\over\sqrt{\pi}}\,,\qquad
&G_+(i)={\pi\over\Gamma({1+x\over2})}
 e^{ -{x\over2}(1-\ln{x\over2}) }\,.
\label{leftconst}
\eea
So the asymptotic series of the free energy density can be simply written as
\bee \label{bosfreen}
\delta{f}_{\rm L}(h)=-{ h^2 \over\pi}\,\,\left(t+(1-{p\over2})\ln t +
A+\ldots\right)\,,
\ee
where the constant $A$,
which enters to our final result for the ratio $m/\Lambda$ is found to be
\bee \label{a}
A={5\over2}\ln2-\ln\pi-1+\ln\Gamma\left({1+x\over2}\right)+{x\over2}(1-\ln2)\,.
\ee

At this point we would like to remark that although the $p\to\infty$
limit in the $S$-matrix 
is smooth (c.f.\ Eq.\ (\ref{pkernel}))
this is not the case in the final results for the free energy, i.e.\ in Eqs.\
(\ref{ferfreen}), (\ref{bosfreen}). This is not so surprising since
Eqs.\ (\ref{ferfreen}), (\ref{bosfreen}) have been obtained  
as asymptotic expansions valid for $h/m\to\infty$ for fixed $p$,
and the
lack of a smooth $p\to\infty$ limit of the coefficients shows 
that the $h/m\to\infty$ and the $p\to\infty$ limits cannot
be simply interchanged in these asymptotic expansions.

Next it might be worth pointing out the following (somewhat puzzling) fact.
In Eq.\ (\ref{ferfreen})
the ratio of the
coefficients of the $\ln t/t^2$ term,
 $-p^2/4$, and the $-1/t$ term,
$p/2$,  is $r_{\scriptscriptstyle\rm D}=-p/2$, while in Eq.\ (\ref{bosfreen})
the ratio of the coefficients of the $\ln t$ term, $1-p/2$,
and that of the $t$ term  is $r_{\scriptscriptstyle\rm L}=1-p/2$.
This appears to be a serious problem since on general grounds
in a theory with a coupling, $\lambda$,
the perturbative result
for a fermionic type free energy is expected to be of the form
\bee
\delta{f}_{\rm Fer}(h)=-{ h^2 f_0\over\pi}\,\,\left(1-f_1{\bar\lambda(h)\over2\pi}+
{\cal O}(\bar\lambda^2)\right)\,,
\ee
while in the same theory for a bosonic case it is expected to be given by
\bee
\delta{f}_{\rm Bos}(h)=-{ h^2 b_0\over\pi}\,\,\left({1\over\bar\lambda(h)}+{\rm
const.}+ {\cal O}(\bar\lambda)\right)\,,
\ee
where $f_0$, $f_1$, $b_0$ are constants and $\bar\lambda(h)$ is the running
coupling.
Since from the renormalization group 
\bee
{1\over\bar\lambda(h)}=\beta_0\ln{h\over\Lambda}+
{\beta_1\over\beta_0}\ln\ln{h\over\Lambda}+{\cal O}({1\over h/\Lambda})\,,
\ee
one would expect $r_{\rm Fer}=r_{\rm Bos}=\beta_1/\beta_0^2$.
As in the present case
$r_{\scriptscriptstyle\rm D}\ne r_{\scriptscriptstyle\rm L}$, it seems
 impossible to match Eq.\ (\ref{ferfreen})
and Eq.\ (\ref{bosfreen}) {\sl simultaneously} with the 
corresponding perturbative expansion.
Fortunately this turns out to be only an apparent paradox which is resolved,
however, in a not entirely obvious way (see Section 3).

Finally we note that the leading term of
$\delta{f}_{\rm L}(h)/h^2$ diverges as $\ln(h/m)$ which can be interpreted
that either the underlying UV current algebra does not exist
or that if it could be defined its level is divergent. 

\subabschnitt{2.4 The left current for $p<1$}
In the previous subsection it has been assumed that the parameter, 
$x=p-1>0$.
Although in the $S$-matrix any $p\geq 0$ is allowed 
in Eq.\ (\ref{leftkernel}) the range of $p$ is, however, restricted to $p>1$.
This restriction follows from the fact that in Eq.\ (\ref{leftkernel}) 
$\lim_{\om\to\infty}\hat{K}_{\rm L}(\om)=1$ for $0\leq p<1$,
that is for $p<1$ $\hat{K}_{\rm L}(\om)$ ceases to be the Fourier 
transform of a function. 
Then clearly one cannot just simply analytically continue in $p$ 
the result for the free energy, (\ref{bosfreen}), for $p<1$.

In order to clarify what happens when $p$
becomes smaller than one, we recall that then in the Sine-Gordon (SG) 
theory bound states (breathers) appear
in the spectrum (for $p<1$ one enters simply the attractive regime).
The SG mass spectrum is given by the well known formula:
\bee\label{SGspectrum}
m_r=2m\sin {\pi pr\over2}\,,\qquad r=1,2\ldots <{1\over p}\,,
\ee
where $m$ denotes the mass of the SG kinks.
As the $S$ matrix of the present model, Eq.\ (\ref{Smatrix}), contains
the SG $S$ matrix as a factor the spectrum then changes accordingly,
i.e.\ breathers also appear (with apropriate multiplicity) in the deformed
NLS model. 
(We remark that $m$ is
also the mass of the SU(2) doublets, of course).

The appearence of these new particles implies that
as soon as $p$ becomes smaller than $1$, 
 the vacuum changes as then 
it becomes energetically favourable for the $r=1$ charge (2,0) 
breather to condense. 
This charge (2,0) breather can be interpreted as the bound state of
the (1,1) and (1,-1) particles (kinks).
In fact as this particle has 
the highest charge/mass ratio, we can again assume (as in Subsection 2.1)
that the true vacuum consists {\sl only} of the condensate of the $r=1$
charge (2,0) breather.
Then one needs the scattering phase, $\delta^{(2,0)}(\theta)$, 
of the (2,0) breathers in order to compute $\delta{f}_{\rm L}(h)$
for the $p<1$ case.

The calculation of $\delta^{(2,0)}(\theta)$ amounts to a simple application
of the well known bootstrap-fusion method \cite{Karowski}. 
In Appendix B we summarized this method and calculated the breather-breather
phase shift, which is given by Eq.\ (\ref{SBB}). Using Eq.\ (\ref{SGbreather-s})
the Fourier transform of the pertinent kernel is found to be
\bee\label{bbkernel}
 1-\hat{K}_{\rm B}(\om)=2\tanh{\pi\vert\om\vert\over2}
\cosh{\pi\om\vert x\vert\over2}\,
 e^{-{\pi\vert\om\vert\vert x\vert\over2}}\,,
\ee
where $\vert x\vert=\vert p-1\vert=1-p$.
The functions appearing in the decomposition of
$ 1-\hat{K}_{\rm B}(\om)$ are now given as
\bee \label{bbgamma}
 G_+(\om)= {1\over\sqrt{-i\om}}{1\over\sqrt{\pi}}
 { \Gamma(1-{i\om\over2})\Gamma(\h-{i\om \vert x\vert\over2})\over
\Gamma(\h -{i\om\over2}) }\,
e^{ -{i \vert x\vert\om\over2}\left(1-\ln(-{i\vert x\vert\om\over2})
\right) }\,, 
\ee 
and again $G_-(\omega)=G_+(-\omega)$.

For the kernel (\ref{bbkernel}) the constants in (\ref{leftfreen})
are found to be:
\bea
&a={\vert x\vert\over2}\,,\qquad
&b=\h \left[\vert x\vert\ln\vert x\vert +\vert x\vert
(\ln2-1-\Gamma'(1))-2\ln2\right]\,,\nonumber\\
&k={1\over\sqrt{\pi}}\,,\qquad
&G_+(i)=\h \Gamma\left({1+\vert x\vert\over2}\right)
 e^{ {\vert x\vert\over2}(1-\ln{\vert x\vert\over2}) }\,.
\label{bbconst}
\eea
So the final result for the asymptotic expansion of the free energy density
for the $r=1$ charge (2,0) breather condensate can be written as 
\bee \label{bbfreen}
\delta{f}_{\rm B}(h_{\rm B})=-{ h_{\rm B}^2 \over4\pi}\,\,
\left(\tau+(1-{p\over2})\ln\tau+
A_{\rm B}+\ldots\right)\,,
\ee
where $\tau=\ln h_{\rm B}/m_1$
and the constant $A_{\rm B}$ is given as
\bee \label{bba}
A_{\rm B}=3\ln2-{3\over2}-\ln\pi+\ln\Gamma\left({p\over2}\right)
+{p\over2}(1-\ln2)+\ln\left(\sin{\pi p\over2}\right)\,.
\ee
It is now easy to verify that 
expressing the breather mass, $m_1$, through the kink mass, $m$,
using the SG spectrum formula Eq.\ (\ref{SGspectrum}) and taking into
account that the charge of the breather is {\sl twice} that of the kink,
i.e.\ $h_{\rm B}=2h$, the free energy density of the breather condensate,
$\delta{f}_{\rm B}(h_{\rm B},m_1)$, 
becomes {\sl identical} to that of the charge $(1,\pm1)$ kink mixture,
$\delta{f}_{\rm L}(h,m)$ analytically continued for negative values of $x$.
This is very reassuring as the result
for $\delta{f}_{\rm L}(h,\Lambda)$ from perturbation theory
turns out to be {\sl continuous} for the full range $0\leq p<\infty$,
which is certainly to be expected.

\subabschnitt{2.5 The O(3) model as the $p\to0$ limit.}
In this last subsection we discuss the $p\to0$ limit and recapitulate
some arguments advanced a long time ago by Wiegmann \cite{Wiegmann}
that the SU(2)$_{\rm L}\times$U(1)$_{\rm R}$ invariant theory (\ref{Lag0})
in this
limit becomes equivalent to the O(3) $\sigma$-model and a decoupled
free massive particle. First of all from Eq.\ (\ref{SGspectrum})
one sees that $m/m_1\to\infty$ as $p\to 0$ i.e.\ the kinks
`disappear' from the spectrum. Furthermore the higher breathers
with $r>1$ also `disappear' from the spectrum, but for a different
reason.
The binding energy of the $r>1$ breathers 
(which can be considered as bound states of the $r=1$ ones)
tends to zero giving rise to zero energy bound states.
Therefore the limiting spectrum consists of 4 massive particles,
transforming as $(3_{\rm L}+1_{\rm L}, 1_{\rm R})$ under
SU(2)$_{\rm L}\times$U(1)$_{\rm R}$.
As already claimed in Ref.\ \cite{Wiegmann} and confirmed in
Ref.\ \cite{Kirillov} in the $p\to0$ limit the $S$-matrix of
these remaining $r=1$ kinks becomes effectively $S_{\rm O(3)}\otimes 1$
i.e.\ the tensor product of the (minimal) $S$ matrix of the O(3) model
proposed by the Zamolodchikov brothers
\cite{Zamol} with the $S$-matrix of a free boson,
consistently with this somewhat heuristic argument.
As this is a highly non-trivial limit it is clearly desirable
to perform as many quantitative checks on it as possible.
In the following we add yet another piece of evidence corroborating
the arguments of Wiegmann, by showing that the $p\to0$ limit of the
$m_1/\Lambda_{\MSb}$ ratio agrees precisely with the one 
calculated directly in the O(3) model \cite{history}.
 
%\newpage
\setcounter{section}{3} \setcounter{equation}{0}
\abschnitt{3. Perturbative Calculations}
 
In this section we calculate the ground state energy of the system
in the presence of external fields in perturbation theory.
As in the previous section, we shall consider fields coupled
to the Noether charges corresponding to the 
U(1)$_{\rm L}\times$U(1)$_{\rm R}$ transformation
\bee
\delta G=i\epsilon_L\sigma^3G-i\epsilon_RG\sigma^3\,.
\ee
To calculate the ground state energy
one has to gauge the symmetries in the Euclidean field theory formulation
and consider the case of constant, imaginary gauge fields \cite{history}.
The corresponding covariant derivative $D_\mu$ is given by
\bee
D_2G=\partial_2G+h_L\sigma^3G-h_RG\sigma^3\,,\qquad\qquad
D_1G=\partial_1G\,.
\label{cov}
\ee 
Putting (\ref{cov}) into (\ref{Lag0}) we write the gauged Lagrangian as:
\bee
{\cal L}={\cal L}_0+{\cal L}_1+{\cal L}_2\,,
\label{Lag}
\ee
where ${\cal L}_1$ and ${\cal L}_2$ denote the terms linear and
quadratic in the external fields, respectively.

To actually calculate the free energy in the presence of the external field,
we take the master formula
\bee
{\cal F}(h)=\delta f(h)=f(h)-f(0)\,,
\label{delF}
\ee
where
\bee
e^{-\int d^nx\,f(h)}=\int{\cal D}G\,{\rm exp}\Big\{-\int d^nx\,({\cal L}_0+
{\cal L}_1+{\cal L}_2)\Big\}\,,
\label{master}
\ee
and expand it in powers of the (bare) coupling  $\lambda_0$. From now 
on the bare lagrangian parameters (couplings)
will be denoted as $\lambda_0$ and $g_0$. 
We have also written the volume element as $d^nx$, where $n=2-\epsilon$
to indicate
that we are going to use dimensional regularization in the perturbative
calculations.
 
In perturbation theory the action has to be expanded around a
(stable) solution of the classical equations of motion. Thus our
next task is to find elements of SU(2), $G_0$, that correspond
to (local) minima of the quadratic part of the gauged Lagrangian,
\bee
{\cal L}_2=-{2h_L^2\over\lambda_0}(1+g_0z^2)-{2h_R^2(1+g_0)\over\lambda_0}+
{4h_Lh_R(1+g_0)z\over\lambda_0}\,,
\label{Lag2}
\ee
where
\bee
z={1\over2}{\rm Tr}\,\Big\{\sigma^3G^{-1}\sigma^3G\Big\}\,,
\ee
which plays the role of the 
potential energy.
In this paper we shall consider
the following two solutions. The first one, called \lq bosonic', (BOS) is
given as:
\bee
G_0={1\over\sqrt{2}}
(1+i\sigma^2)\,,\quad h_R=0\,.
\label{G0BOS}
\ee
The other one, which shall be referred to as \lq fermionic' (FER) is written as 
\bee
G_0=i\sigma^2\,,
\label{G0FER}
\ee
where we also require that $h_L$ and $h_R$ satisfy the inequality
\bee
\Delta=4\big[h_L^2g_0+h_Lh_R(1+g_0)\big]\geq0\,,
\label{ineq2}
\ee
ensuring the non-negativity of the mass term.

The terms \lq bosonic' (resp.\ \lq fermionic') are used
here by anticipating that they actually correspond to the bosonic (resp.\
fermionic) type free energies discussed in the previous section.
The BOS-type solution will be used to calculate the free energy corresponding
to the pure U(1)$_{\rm L}$ charge ($h_R=0$, LEFT) case,
while the FER-type solution is appropriate to
calculate the free energy for both the pure U(1)$_{\rm R}$ charge 
($h_L=0$, RIGHT) and the
\lq diagonal' (DIAG) case, where both $h_L$ and $h_R$ are different from
zero.

We now start to analyse the BOS case first.
The leading, ${\cal O}(\lambda_0^{-1})$, term in perturbation theory is
given by the potential energy, ${\cal L}_2$, at its minimum, which 
is immediately seen to be ($z_0=0$, $h_R=0$)
\bee
{\cal F}^{(-1)}={\cal L}^{(-1)}=-{2h_L^2\over\lambda_0}
\label{FclassBOS}\,.
\ee
 
Then we turn to the calculation of the next, 1-loop, term in perturbation
theory. The calculation is elementary but for completeness
we give some details in Appendix C. The general 1-loop contribution is given
by Eq.\ (\ref{F0}), which is a rather complicated function of three parameters.
It simplifies somewhat in the two cases we are
interested in. In the present, BOS, case after using (\ref{paraBOS}),
we obtain
\bee
{\cal F}^{(0)}={4h_L^2\over n}\int{d^np\over(2\pi)^n}\,
{(1+g_0)\,p_2^2-g_0p^2\over
(p^2)^2-4g_0h_L^2p^2+4(1+g_0)h_L^2p_2^2}\,.
\label{F0BOS}
\ee
Because of the explicit dependence of the integrand
on the second component of the momentum we have to modify
the usual rules of dimensional regularization by the following.
In all Feynman integrals we perform the one-dimensional integral
over $p_2$ first, followed by the integration over the
remaining momentum components, continued to $n-1$ dimensions, i.e.\ in
the present paper
\bee\label{dimrule}
\int{d^np\over(2\pi)^n}:=\int{d^{n-1}p\over(2\pi)^{n-1}}\int
\limits_{-\infty}^\infty{dp_2\over(2\pi)}\,.
\ee
It is possible to calculate the integral (\ref{F0BOS}) exactly \cite{Karp}.
For our purposes, however, it is sufficient to
calculate its divergent part (the part that becomes singular in the
$\epsilon\rightarrow0$ limit) and to show that ${\cal F}^{(0)}$ must be of
the form
\bee
{\cal F}^{(0)}=h_L^n\,\Bigg\{{1-g_0\over2\pi\epsilon}+W(g_0)\Bigg\}\,,
\label{F0BOS1}
\ee
where $W(g_0)$ is a (complicated)
finite function of its argument. Fortunately all we shall
need is the value of $W(g_0)$ at the special point $g_0=-1$.
For this very special $g_0$
(\ref{F0BOS}) simplifies enormously:
\bee
{\cal F}^{(0)}\Big\vert_{g_0=-1}
={4h_L^2\over n}\int{d^np\over(2\pi)^n}\,{1\over
p^2+4h_L^2}\,.
\label{F0BOS2}
\ee
Calculating (\ref{F0BOS2}) and comparing the result to (\ref{F0BOS1}) yields
\bee
W(-1)={1\over2\pi}\,\big[1+\Gamma^\prime(1)+{\rm ln}\,\pi\big]\,.
\ee
Having obtained the bare 1-loop result,
we now turn to renormalization
and renormalization group (RG) improvements. The methods are well-known but
we summarized the details of the calculation in Appendix D to make this
paper relatively self-contained.
 
The RG improved perturbative result gives
the asymptotic expansion of the free energy for large values of the external 
fields. Writing the sum of the classical
contribution (\ref{FclassBOS}) and the one-loop correction (\ref{F0BOS1})
and using definitions (\ref{Z1}), (\ref{Z2}) we obtain the renormalized form
\bee
{\cal F}(h)=-{2h^2\over\lambda}-{(1-g)h^2\over4\pi}\,\ln{h^2\over\mu^2}+
h^2W(g)+{\cal O}(\lambda)\,,
\label{FrenBOS}
\ee
where $h=h_L$. 
Expressing the
result in terms of the single effective coupling, $\eff$, defined by
Eq.\ (\ref{eff}) we get
\bee
{\cal F}(h)=-2h^2\Bigg\{{1\over \eff}+{1\over2\pi}\Big(\ln2-{1\over2}\Big)
+\cdots\Bigg\}\,.
\label{FeffBOS}
\ee
At this point we remark that we have identified the RG invariant parameter
defined in terms of the running couplings $\bar g(t)$
and $\bar\lambda(t)$ by Eq.\ (\ref{g}) 
with the `deformation parameter' $p$ appearing in the $S$ matrix 
(\ref{Smatrix}).
Finally expanding (\ref{FeffBOS}) with the help of (\ref{e1}) the free
energy in the BOS case can be expressed as
\bee
{\cal F}(h)=-{h^2\over\pi}\Bigg\{s+\Big(1-{p\over2}\Big)\ln s+
\Big(\ln2-{1\over2}\Big)+\cdots\Bigg\}\,.
\label{FexpBOS}
\ee

Next we turn to the computation in the FER case. The leading (classical)
term of the free energy is then  
\bee
{\cal F}^{(-1)}={\cal L}^{(-1)}=-{2(1+g_0)\over\lambda_0}\,(h_L+h_R)^2\,.
\label{FclassFER}
\ee
Note that ${\cal F}^{(-1)}$  for the diagonal
charge depends {\sl only} on the sum $h_L+h_R$.
This is consistent with the results  
of the TBA analysis found in Section 2.\ for the DIAG case.  
In the present (FER) case (\ref{F0}) becomes
\bee
{\cal F}^{(0)}={2\over n}\int{d^np\over(2\pi)^n}\,{\rm Re}\,
{\Delta-i\omega p_2\over
p^2+\Delta-2i\omega p_2}\,,
\label{F0FER}
\ee
where
\bee
\omega=h_L(1-g_0)-h_R(1+g_0)\,.
\label{omega}
\ee
Evaluating Eq. (\ref{F0FER}) using the modified dimensional scheme
(c.f.\ (\ref{dimrule})) yields:
\bee
{\cal F}^{(0)}=(1+g_0)^n\,(h_L+h_R)^n\Bigg\{{1\over2\pi\epsilon}+{1\over4\pi}
\Big[1+\Gamma^\prime(1)+{\rm ln}\,4\pi\Big]\Bigg\}\,.
\label{F0FER1}
\ee
It is gratifying to find 
that the one loop term (\ref{F0FER1}) also depends {\sl only} on the sum 
$h_L+h_R$, just like
the classical one (\ref{FclassFER}). This is not an accident.
It is in fact absolutely essential that ${\cal F}(h_L,h_R)$ 
be a function of $h_L+h_R$ in order to
be able to match the perturbative result with the one found by the TBA
method for the DIAG case in Subsection 2.1.
As shown in Appendix C, in the FER case the free energy does depend only
on the sum $h_L+h_R$ to {\sl all orders} in perturbation theory.
Moreover it is easy to see that
the free energy in the FER case must be of the form
\bea
{\cal F}(h)&=& -{2h^2(1+g_0)\over\lambda_0}+{M^2\over\lambda_0}\,
\sum_{L=1}^\infty\, F_L(g_0)\,\Bigg({\lambda_0\over M^\epsilon}\Bigg)^L
\label{FallordI}\\
&=&-2h^2\,\Bigg\{{1+g_0\over\lambda_0}-{(1+g_0)^2\over2\lambda_0}\,
\sum_{L=1}^\infty\,F_L(g_0)\,(\lambda_0 M^{-\epsilon})^L\Bigg\}\,,
\label{FallordII}\\
\nonumber
\eea
where 
\bee
M=(1+g_0)\,(h_L+h_R)\,,
\label{M}
\ee
and the summation is over the number of loops, $L$.
(\ref{FallordI}) is nothing but dimensional analysis based on the
fact that the only dimensionful parameter available is $M$,
which serves as an infrared cutoff in our perturbative
computations. The second form (\ref{FallordII}) illustrates the fact
that all higher order terms are proportional to $(1+g_0)^2$, which fact plays an
important role in our calculations.

Adding now the classical and the one-loop terms,
(\ref{FclassFER}) and (\ref{F0FER1}) and expressing the result in terms of
renormalized quantities we obtain:
\bee
{\cal F}(h)=-2h^2\Bigg\{{1+g\over\lambda}+{(1+g)^2\over8\pi}\Bigg[
\ln\Bigg({(1+g)^2h^2\over\mu^2}\Bigg)-1-2\gamma\Bigg)
+{\cal O}(\lambda)\Bigg\}\,,
\label{FrenFER}
\ee
where $h=h_L+h_R$ and the constant $\gamma$ is defined as
\bee
\gamma={\scriptstyle{1\over2}}\,\Gamma^\prime(1)+\ln\sqrt{4\pi}\,.
\label{smallgamma}
\ee
As we need the expansion of ${\cal F}(h)$ up to
${\cal O}(\eff^2)$ in the effective coupling in this case,
this would necessitate a three-loop computation at first sight. 
In fact we obtain all the ${\cal O}(\eff^2)$ terms from the one loop result
(\ref{FrenFER}) alone! This `mini miracle' is due to the
$(1+g)^2$ factor in front of the one-loop term
together with the fact that $(1+\bar g)^2={\cal O}(\eff^2)$, implying that
the one-loop term is already ${\cal O}(\eff^2)$.
Moreover as it follows from Eq.\ (\ref{FallordII}) the contribution of the 
higher order terms will be
at least ${\cal O}(\eff^3)$ hence they can be safely ignored. 
Putting then everything together we find
\bee
{\cal F}(h)=-{ph^2\over\pi}\Bigg\{1-{p\over4\pi}\eff+
{p^2\over32\pi^2}\eff^2+
{p\eff^2\over8\pi^2}\Bigg[\ln p+\ln\Big({\eff\over2\pi}\Big)\Bigg]+
{\cal O}(\eff^3)\Bigg\}\,.
\label{FeffFER}
\ee
Note the non-analytic contribution, $\ln \eff$, in the last term of
(\ref{FeffFER}).
This comes ultimately from the $g_0$-dependence in (\ref{M}) of $M$, which
plays the role of the infrared cutoff in our calculations. The presence of
this non-analytic term in  $\eff$
explains why the coefficient of the $\ln s$ term differs
for the BOS and the FER cases. 
Indeed using Eq.\ (\ref{e1}) we find that the large $s$ expansion
in the FER case is finally given as
\bee
{\cal F}(h)=-{ph^2\over\pi}\Bigg\{1-{p\over2s}-{p^2\over4s^2}\ln s+
{p\over2s^2}\Bigg[\ln p+{p\over4}\Bigg]+\cdots\Bigg\}\,.
\label{FexpFER}
\ee
 
Having calculated the asymptotic form of the free energy both in
perturbation theory and with the TBA method,
by comparing them (with $t=s-\ln(m/\Lambda_{\overline{\rm MS}})$)
we obtain the relation between the mass of the
doublet particles, $m$, and $\Lambda_{\overline{\rm MS}}$. 
In both the BOS resp.\ FER case the comparison of Eqs.\ (\ref{bosfreen})
and (\ref{FexpBOS}) resp.\ (\ref{ferfreen}) and (\ref{FexpFER}) leads
to the result (\ref{MperLambda}) as already announced in the Introduction.

We emphasize that it is already a very nontrivial check on the overall
consistency of our assumptions that all the expansion coefficients,
for both the BOS and the FER cases, agree. A further, very stringent
consistency check is that the $m/\Lambda$ ratio obtained in the BOS
case is exactly the same as the one obtained in the FER case.
Finally by comparing the free energy of the breather condensate,
computed from the TBA method, Eq.\ (\ref{bbfreen}),
with the perturbative one, Eq.\ (\ref{FexpBOS}),
one finds  
\bee
{m_1\over\Lambda_{{\overline{{\rm MS}}}}}=2^{3-{p\over2}}\,
e^{{p\over2}-1}\,
\Gamma\big(1+{p\over2}\big){2\sin{\pi p\over2}\over\pi p}\,.
\label{M1perLambda}
\ee
By recalling that the mass of the lowest lying ($r=1$) breather  
is $m_1=2m\sin{\pi p\over2}$ one immediately obtains 
from Eq.\ (\ref{M1perLambda}) the $m/\Lambda$ ratio in Eq.\ (\ref{MperLambda}).
This is clearly a further nontrivial check on our result
and on the mutual consistency of the hypothesises made in the course of
the calculation.

The ratio (\ref{MperLambda}) diverges in the $p\rightarrow\infty$
limit. This illustrates the fact, discussed in Subsection 2.3, that 
it is not possible to recover the asymptotic expansion of the
ground state energy of
the $O(4)$ model from the generic case in this limit.
Furthermore $m/\Lambda_{\overline{\rm MS}}$
also diverges as $p\rightarrow0$ consistently with
the Wiegmann scenario as discussed in Subsection 2.5, while
$$\lim_{p\to0}{m_1\over\Lambda_{{\overline{{\rm MS}}}}}={8\over e}\,,$$
which exactly reproduces the $m/\Lambda$ ratio 
of the O(3) model first computed 
in Ref.\ \cite{history}. 
In our view this beautiful agreement certainly yields some more 
quantitative support as to the correctness 
of the `conventional wisdom'.

\setcounter{section}{4} \setcounter{equation}{0}
\abschnitt{4. Lattice regularization}
 
In this section we propose a lattice action for the deformed
principal model and calculate the free energy in this lattice
version of the model. This is then used to calculate the ratio
of the $\Lambda$-parameters of the dimensionally regulated
theory and the lattice version.
 
First we rewrite the Lagrangian (\ref{Lag0}) in the form:
\bee
{\cal L}_0=(1+g){\cal L}^{O(4)}-g{\cal L}^{O(3)}\,,
\label{Lag34}
\ee
exhibiting that it interpolates between the $O(3)$ and $O(4)$
nonlinear $\sigma$-models. In fact
\bee
{\cal L}^{O(4)}={1\over\lambda}\,{\rm tr}\Big\{
\partial_\mu G^{-1}\partial_\mu G\Big\}\,,
\label{Lag4}
\ee
is the familiar action of the SU(2) principal chiral (or O(4)) nonlinear
$\sigma$-model and
\bee
{\cal L}^{O(3)}={1\over\lambda}\,
\partial_\mu \Sigma^a \partial_\mu \Sigma^a\,,
\label{Lag3}
\ee
is the Lagrangian of an O(3) symmetric nonlinear $\sigma$-model with
a composite $O(3)$ field  
\bee
\Sigma^a={1\over2}\,{\rm tr}\Big\{
G^{-1}\sigma^a G\sigma^3\Big\}\,.
\label{Sigma}
\ee
 
Motivated by this form of the Lagrangian we
take as our lattice action the following linear combination.
\bee
S_{\scriptscriptstyle {\rm Lattice}}=
(1+\tilde{g}_0)\,S^{O(4)}-\tilde{g}_0\,S^{O(3)}\,.
\label{SLatt}
\ee
Here
\bee
S^{O(4)}={1\over\tilde{\lambda}_0}\,\sum_{x,\mu}\,{\rm tr}\,\Big\{
2-G^{-1}(x)G(x+\hat\mu)-G^{-1}(x+\hat\mu)G(x)\Big\}\,,
\label{S4}
\ee
is the standard O(4) lattice action in terms of the SU(2)
principal model variable $G(x)$ associated to the lattice site
$x$. (As usual, the lattice spacing $a$ is taken to be
unity and $\hat\mu$ denotes the unit vector in the $\mu$
direction. Moreover, we have denoted the bare lattice couplings by
$\tilde{\lambda}_0$ and $\tilde{g}_0$ to distinguish them from their
counterparts in the dimensional scheme.) Similarly
\bea
S^{O(3)}&=&{1\over\tilde{\lambda}_0}\,\sum_{x,\mu}\,\Big\{
1-\Sigma^{a}(x)\Sigma^a(x+\hat\mu)\Big\}\label{S3}\\
&=&{1\over2\tilde{\lambda}_0}\,\sum_{x,\mu}\,{\rm tr}\,\Big\{
G(x)\sigma^3G^{-1}(x)G(x+\hat\mu)\sigma^3G^{-1}(x+\hat\mu)-1\Big\}\,,
\nonumber\\
\nonumber
\eea
is the standard O(3) lattice action in terms of the composite
field (\ref{Sigma}).
 
For simplicity we shall consider here the free energy
associated to the $U(1)_{\rm L}$ charge only.
This corresponds to the BOS case of Section 3  
(i.e.\ $h_L=h$, $h_R=0$).
An apropriate gauging of the Lagrangian (\ref{SLatt}) and taking imaginary
gauge fields leads to:
\bea
S_{\scriptscriptstyle {\rm gauged}}&=&
{1+\tilde{g}_0\over\tilde{\lambda}_0}\,\sum_{x,\mu}\,{\rm
tr}\,\Big\{ 2-G^{-1}(x)e^{-h_\mu\sigma^3}
G(x+\hat\mu)-G^{-1}(x+\hat\mu)e^{h_\mu\sigma^3}G(x)\Big\}
\nonumber\\
&+&{\tilde{g}_0\over2\tilde{\lambda}_0}\,\sum_{x,\mu}\,{\rm tr}\,\Big\{
G(x)\sigma^3G^{-1}(x)e^{-h_\mu\sigma^3}
G(x+\hat\mu)\sigma^3G^{-1}(x+\hat\mu)e^{h_\mu\sigma^3}
-1\Big\}\,,\nonumber\\
\label{Sgauged}
\eea
with $h_\mu=h\delta_{\mu2}$.
 
The perturbative calculation of the free energy on the lattice
is now completely analogous to the one in the dimensional scheme.
The leading contribution is given again by the minimum of the gauged 
action for constant fields. (\ref{Sgauged}) also takes its minimum at
the group element (\ref{G0BOS}) and is given by
\bee
{\cal F}_L^{(-1)}=-{4(1+\tilde{g}_0)\over\tilde{\lambda}_0a^2}\,
\Big(\cosh ha-1\Big)
+{\tilde{g}_0\over\tilde{\lambda}_0a^2}\,\Big(\cosh 2ha-1\Big)\,,
\label{FclassLatt}
\ee
where the $a$-dependence has been reconstructed. In the continuum
limit, $a\rightarrow0$, (\ref{FclassLatt}) reproduces (\ref{FclassBOS}).

The 1-loop contribution is half the logarithmic determinant of the
operator corresponding to the quadratic term in the expansion of the
lattice action around the group element (\ref{G0BOS}). We have not tried
to calculate the 1-loop determinant for generic $\tilde{g}_0$, because 
for our purposes it is sufficient to show that it must be of the form
\bee
{\cal F}_L^{(0)}=-h^2\Bigg\{{1-\tilde{g}_0\over2\pi}\,\ln ha
+W_L(\tilde{g}_0)+{\cal O}(a^2)\Bigg\}\,.
\label{F0Latt}
\ee
$W_L(\tilde{g}_0)$ similarly 
to its dimensional
counterpart, is a complicated function of its argument but all we are 
going to need is its value at the $O(3)$ point $\tilde{g}_0=-1$, where it is
relatively easy to evaluate. We find
\bee
W_L(-1)=-{1\over2\pi}-{3\over2\pi}\ln2-{5\over8}\,.
\label{W1Latt}
\ee
 
We now proceed, as before, from the bare results (\ref{FclassLatt})
and (\ref{F0Latt}) via the renormalized ones to the RG improved
1-loop formula 
\bee
{\cal F}_L(h)=-{h^2\over\pi}\Bigg\{\ln{h\over\Lambda_L}+
\Big(1-{p\over2}\Big)\ln\ln{h\over\Lambda_L}-{1\over2}-
{3\over2}\ln2-{5\pi\over8}\Bigg\}+\dots\,,
\label{FexpLatt}
\ee
where we have used the results of Appendix D. Comparing (\ref{FexpBOS})
and (\ref{FexpLatt}) gives the ratio of the $\Lambda$-parameters of the
two regularization schemes:
\bee
{\Lambda_{\overline{\rm MS}}\over\Lambda_L}=\sqrt{32}\,
e^{{5\pi\over8}}\,.
\label{Lambdaratio}
\ee
Note that the ratio (\ref{Lambdaratio}) is independent of the RG-invariant
parameter $p$, but is nevertheless different from the corresponding ratio
of the standard $O(3)$ model, which is given by a formula similar to
(\ref{Lambdaratio}), but with an exponent that is $\pi/2$ instead of
$5\pi/8$. 
This means that our lattice regularization is different from the
standard one in the $O(3)$ limit. A possible explanation
for this might have been that the difference is due to the
compositeness of our $O(3)$ field (\ref{Sigma}). But
by a simple computation we find the standard $\Lambda_L$ for
$\tilde g_0=-1$ if the first term in (\ref{SLatt}) is simply omitted.
To clarify this apparent paradox we note that the difference
is due to the following term in the quadratic piece of the gauged
lattice action.
\bee
S_\Phi={4(1+\tilde{g}_0)\over\tilde{\lambda}_0}\,\sum_{x,\mu}\,
\cosh h_\mu\,\Phi(x)\Big[\Phi(x)-\Phi(x+\hat\mu)\Big]\,.
\label{SPhi}
\ee
(\ref{SPhi}) is, of course, absent if $\tilde{g}_0=-1$. This is the
case in the standard lattice version of the $O(3)$ model. If, however,
$\tilde{g}_0\not=-1$, then the coupling dependence can be scaled out of
(\ref{SPhi}). Close to the continuum limit, after rescaling, (\ref{SPhi})
corresponds to the continuum Lagrangian
\bee
{\cal L}_\phi={1\over2}\,\partial_\mu\phi\,\partial_\mu\phi+
{1\over4}\,h^2a^2\,(\partial_2\phi)^2+{\cal O}(h^4a^4)\,,
\label{Lagphi}
\ee
from which one would  naively conclude that this field is
decoupled from the external field $h$ in the continuum limit. This
is, however, not correct since the second term in (\ref{Lagphi})
does in fact contribute to the continuum free energy, because of 
quadratic divergences in lattice perturbation theory.
This contribution is
$h^2/8$ leading to the modified exponent in (\ref{Lambdaratio}).
We expect to find the same modified $\Lambda_L/\Lambda_
{{\overline{{\rm MS}}}}$ ratio by comparing the 4-point function
computed in the two regularization schemes.

For completeness we also computed the 
$\Lambda_{{\overline{{\rm MS}}}}/\Lambda_L$ ratio in the $O(4)$
symmetric case $g=0$. In this special case the function defined
in (\ref{F0BOS1}) is relatively easy to calculate. In the
dimensional regularization scheme
\bee
W(0)={1+2\gamma\over4\pi}\,,
\label{W0dimreg}
\ee
whereas in the analogous calculation for lattice regularization we find
\bee
W_L(0)={1\over4\pi}\Big[1+\ln32+{\pi\over2}\Big]\,.
\label{W0lattice}
\ee
Using (\ref{W0dimreg}) and (\ref{W0lattice}) we obtain the
$\Lambda_{{\overline{{\rm MS}}}}/\Lambda_L$ ratio
\bee
{\Lambda_{{\overline{{\rm MS}}}}\over\Lambda_L}=
\sqrt{32}\,e^{{\pi\over4}}\,,
\ee
in agreement with the well-known result. We note that what is new
in our calculation is that we considered the external field
coupled to $Q_L$ here instead of the diagonal charge considered
previously. The $\Lambda_{{\overline{{\rm MS}}}}/\Lambda_L$ ratio is the
same of course, as it should.

We think that it would be extremely interesting to perform Monte Carlo
simulations using the lattice action (\ref{SLatt}). This would provide us
with an independent way of determining the ratio of the mass of the
physical particles and the lattice $\Lambda$ parameter and  also with an
interesting example of a lattice field theory model with more than one 
relevant couplings.

%\newpage 

\setcounter{section}{5} \setcounter{equation}{0}
\abschnitt{5. Summary and Conclusions}
 
In this paper we have computed the ground state energies
of the deformed principal model in the presence of external
fields for all three possible orientations of the external field.
We have found a complete consistency between perturbation theory
and the results from the TBA  method  based on the proposed S-matrix 
of the model.
In all three cases the leading and the first few subleading
coefficients of the large field expansion for the ground state
energy are identical once we calculate the $m/\Lambda$
ratio using one of the expansions. This beautiful
agreement is absolutely nontrivial since it can only be achieved
after identifying an unusual non-analytic term in the
perturbative expansion for the fermionic case. Such non-analytic
terms can only occur in perturbation theory in the presence of several
couplings. To our knowledge our example is the first one
where the results of the TBA program is compared to
the results of asymptotically free perturbation theory
with several coupling constants.
 
Actually the presence of the running deformation parameter,
$\bar g$, made the computations much easier as compared to
a purely fermionic model. In a 
fermionic model, generically a three-loop calculation is
necessary to obtain the $m/\Lambda$ ratio, whereas in the
deformed model here, it has been sufficient to calculate 
only to  one-loop order! 
This nice simplification is due to the fact
that all higher terms are 
proportional to $(\bar g+1)^2$, and since for large fields
$\bar g\rightarrow-1$, they are effectively supressed.
 
We have also outlined a proof to
all orders of perturbation theory that
the free energies for the two apparently different fermionic
cases (DIAG and RIGHT) 
actually have the same form as functions of the effective
chemical potentials. 
We have no explanation for this
fact but it is yet another very nontrivial consistency check
since the TBA integral equations are {\sl identical} for the two
cases.
 
It is also quite remarkable that the agreement between perturbation
theory and the TBA method is valid for all values of the
RG invariant parameter, $p=2\pi(1+\bar
g)/\bar\lambda$, including the attractive regime
$0<p<1$, where the bootstrap S-matrix indicates the existence
of a complex bound state structure in the spectrum of the model.
We have found that the structure of the ground state is changing
with a decreasing $p$ in the sense that it is always the lowest lying
breather bound state that condenses to the vacuum.
The final
formula for the ground state energy is, however, analytic
for the whole range of $0<p<\infty$ and agrees with the results of PT.
Finally our results are also in perfect agreement with the 
%lends some additional support to the
Wiegmann scenario for the $p\rightarrow0$ limit since
$\lim_{p\to0}m_1/\Lambda=8/e$, ($m_1$ being the mass of the lowest lying
breather), thus we reproduce the $m/\Lambda$ ratio computed previously
in the O(3) model.
What we found here lends also additional support to the empirical rule 
that only the particle with the highest charge/mass ratio condenses
in the ground state. This rule was found to be correct in all cases
considered so far \cite{hollo} and proved to be correct also for the
deformed model in the attractive regime. In view of the
complexity of the spectrum of physical exitations in the attractive
regime it would be extremely
interesting to find an independent explanation of this phenomenon.

We have also proposed a simple lattice version of the SU(2)$\times$U(1)
symmetric model and computed the $\Lambda_\MSb/\Lambda_L$ ratio where
$\Lambda_L$ denotes the lattice lambda
parameter.
%We think it would be worthwile to study
%this lattice model with MC methods not only to also measure 
%the computed $m/\Lambda_L$ ratio, but also because it
%provides a simple model where the effects of the
%presence of two relevant couplings can be studied.
 
An interesting question that requires further study
is the application of the TBA program to calculate
thermodynamical quantities for finite temperature 
and comparing them to the results of perturbative
calculations.

\noindent{\it Acknowledgements}
J. B. gratefully acknowledges a CNRS grant.
This investigation was supported in part by the Hungarian National
Science Fund (OTKA) under T019917 and T 030099.

\renewcommand{\thesection}{\Alph{section}}
\setcounter{section}{1} \setcounter{equation}{0}
\abschnitt{Appendix A. Solution of the TBA integral equation}
 
In the Appendix of Ref.~\cite{FoNiWe} the
solution of the pertinent integral equation determining the free energy
has been obtained for a generic kernel by applying the
generalized Wiener--Hopf technique.
The integral equation is given as follows:
\bee\label{inteq}
\epsilon(\theta)-\int_{-B}^{B}d\theta'K(\theta-\theta')
\epsilon(\theta')=h-m\cosh\theta\,,
\ee
together with the boundary condition $\epsilon(\pm B)=0$.
In terms of the solution of Eq.~(\ref{inteq}) the free energy is
\bee
\delta
f(h)=-{m\over2\pi}\int_{-B}^{B}d\theta\cosh(\theta)\epsilon(\theta)\,.
\ee
We do not go into the details of the solution technique of Eq.~(\ref{inteq})
as they can be found in the literature (e.g.\ Ref.\ \cite{FoNiWe}).
Here we just present the results in a somewhat simplified form which
could prove convenient for future applications.
The crucial step in solving Eq.~(\ref{inteq}) by the generalized
Wiener-Hopf method is
 to split the Fourier transform of the kernel in Eq.\ (\ref{inteq}),
$\hat{K}(\om)$, as
\bee\label{split}
 1-\hat{K}(\om)= \lb G_+(\om) G_-(\om) \rb^{-1}\,,
\ee
where $G_+(\om)$ resp.\ $G_-(\om)$ is analytic in the upper resp.\ lower
half plane.
When the kernel is of the `fermionic' type, i.e.\ $\hat{K}(\om)\ne1$,
one can assume the following expansion of $G_+(i\xi)$ around $\xi=0$ for
$\xi>0$:
\bee\label{GFexpansion}
G_+(i\xi)=\tilde k\exp{\{-\tilde a\xi\ln\xi\}}
\left(1+\tilde b\xi+O(\xi^2)\right)\,.
\ee
Then in the limit of high
density $h/m\gg1$,
the asymptotic series for $\delta f(h)$ can be obtained by a simple iterative
technique. The final result, Eq.\ (A.34) of Ref.\ \cite{FoNiWe},
can be conveniently recast in the following form:
\bee\label{fren1}
 \df=-{h^2 \over 2\pi} \tilde k^2\left\{
1+{\tilde a \over t}-{\tilde a}^2{\ln t \over t^2}
+{\tilde A\over t^2}+\ldots \right\} \,,
\ee
where $t=\ln(h/m)$ and the constant $\tilde A$ is given as
\bee\label{tildeA}
\tilde A=
\tilde a\left( 1-\tilde b-\ln {2\tilde k\over G_+(i)} \right)+
 {\tilde a}^2 \left( {3\over2}+\Gamma'(1)-\ln 2 \right)\,.
\ee 

For `bosonic' type kernels, $\hat{K}(0)=1$
so the splitting
functions $G_+(\om)$, $G_-(\om)$ are singular at $\om=0$.
The solution of the integral equation (\ref{inteq}) is then more complicated
than in the previous (FER-type) case as the corresponding Neumann
series is not uniformly convergent and the solution cannot be obtained
by simple iteration. The solution of Eq.~(\ref{inteq})
for a generic BOS-type kernel in the limit $t=\ln(h/m)\to\infty$
has been given in Ref.~\cite{Balog} up to order of $\ln t/t$.
According to the results of Ref.~\cite{Balog}
for $G_+(i\xi)$  admitting the following expansion around $\xi=0$ ($\xi>0$)
\bee\label{Gexpansion}
G_+(i\xi)={k\over\sqrt{\xi}}\exp{\{-a\xi\ln\xi\}}
\left(1-b\xi+O(\xi^2)\right)\,,
\ee
the asymptotic series for $t\gg1$ of the Legendre transform of ground state
energy density can be expressed in the form
\bee \label{leftfreen}
\delta{f}(h)=-{ h^2 \over4}\,k^2\,\left(t+(a+\h)\ln t +
A+\ldots\right)\,,
\ee
where the constant $A$ is given by
\bee\label{A}
A=\ln\left({\sqrt{2\pi}ke^{-b}\over G_+(i)}\right)-1+
a(\ln8-1-\Gamma'(1))\,.
\ee

\renewcommand{\thesection}{\Alph{section}}
\setcounter{section}{2} \setcounter{equation}{0}
\abschnitt{Appendix B. Bootstrap fusion}
 
In this appendix we calculate the phase shifts corresponding
to the scattering of two bound state particles of charge (2,0)
and of lowest mass $m_1=2m\sin{p\pi\over2}$ in terms of the
scattering phases of the \lq\lq elementary" particles of charge
(1,1) and (1,-1). The procedure is called bootstrap fusion and
being an important part of the whole bootstrap program is discussed
at length in the literature \cite{Karowski,Coleman,Mussardo}.
Therefore we just apply the general framework to
the special case of our interest, but even this could
prove to be useful for other models.
 
We shall assume that the S-matrix describing the scattering
of the \lq\lq elementary" particles of mass $m$
can be decomposed as
\bee
S^{\beta^\prime\alpha^\prime}_{\alpha\beta}(\theta)=
\sum_e\,\varphi^{\beta^\prime\alpha^\prime}_eS_e(\theta)
\varphi^e_{\alpha\beta}\,,
\label{project}
\ee
where the orthogonal projectors satisfy
\bee
\sum_e\varphi^{\alpha\beta}_e
\,\varphi^e_{\alpha^\prime\beta^\prime}=
\delta^\alpha_{\alpha^\prime}\delta^\beta_{\beta^\prime}\,,
\qquad\qquad
\sum_{\alpha,\beta}\varphi^{e^\prime}_{\alpha\beta}
\,\varphi^{\alpha\beta}_e=
\delta^{e^\prime}_e\,.
\label{pro12}
\ee
The scattering phases may have poles in the physical sheet
$0\leq {\rm Im}\theta\leq\pi$ of the form
\bee
S_e(\theta)\approx{iR_{be}\over\theta-iu_b}\,.
\label{poles}
\ee
If the pole in (\ref{poles}) is associated\footnote{
Bound states are not the only possibility to explain the
presence of these, or higher order, poles.
See \cite{Coleman, Mussardo}.} with a bound state of mass
\bee
m_b=2m\cos{u_b\over2}\,,\label{mass}
\ee
then one can use the bootstrap fusion procedure to calculate
the S-matrix of the bound states from the S-matrix of their
constituents. We assume further that the bound state, $b$,
occurs only in one of the chanels (which we shall label
$e=b$) and that the scattering of $b$ is diagonal with all
other (elementary or composite) particles from the spectrum of the
model. Then the scattering of $b$ and $X$ (which can be one
of the constituents or some composite particle) is given by
\bee
S_{bX}(\theta)=\sum_{\alpha,\beta,\tilde X,\tilde\alpha,
\tilde\beta}\,\varphi^b_{\alpha\beta}S^{X\alpha}_{\tilde\alpha
\tilde X}\big(\theta+{iu_b\over2}\big)
S^{\tilde X\beta}_{\tilde\beta
X}\big(\theta-{iu_b\over2}\big)\varphi^{\tilde\alpha\tilde\beta}
_b\,.
\label{fusion}
\ee
 
In the case of the Sine-Gordon model the particles are indexed
by their U(1) charges (which can be 1 or -1) and the four scattering 
chanels are labeled $(0,\bar0,\pm)$. The projectors are 
\cite{Karowski2}
\bea
\varphi^{\alpha\beta}_0&=&\varphi^0_{\alpha\beta}
=\delta^\alpha_1\delta^\beta_1\,,\nonumber\\
\varphi^{\alpha\beta}_{\bar0}&=&\varphi^{\bar0}_{\alpha\beta}
=\delta^\alpha_{-1}\delta^\beta_{-1}\,,\\
\varphi^{\alpha\beta}_\pm&=&\varphi^\pm_{\alpha\beta}
={1\over\sqrt{2}}\big(\delta^\alpha_1\delta^\beta_{-1}
\pm\delta^\alpha_{-1}\delta^\beta_1\big)\,.
\nonumber\\
\nonumber
\eea
The soliton-soliton and antisoliton-antisoliton scattering
phases are
\bee
S_0^{(p)}(\theta)=S_{\bar0}^{(p)}(\theta)=
-e^{i\delta_p(\theta)}\,,
\ee
where
\bee\label{pkernel}
\delta_p(\theta)=\int\limits_{-\infty}^\infty d\omega\,
{\sin(\omega\theta)\over\omega}\hat K_p(\omega)\,,\quad 
\hat K_p(\omega)={ \sinh\left({\pi\omega(p-1)\over2}\right)\over
 2\cosh\left({\pi\omega\over2}\right)
\sinh\left({\pi\omega p\over2}\right) }\,.
\ee
The other two eigenvalues are
\bee
S_+^{(p)}(\theta)=-{\sinh\big({\theta+i\pi\over2p}\big)\over
\sinh\big({\theta-i\pi\over2p}\big)}\,S^{(p)}_0(\theta)\,,
\ee
and
\bee
S_-^{(p)}(\theta)=-{\cosh\big({\theta+i\pi\over2p}\big)\over
\cosh\big({\theta-i\pi\over2p}\big)}\,S^{(p)}_0(\theta)\,.
\ee
The lowest breather has mass $m_b=2m\sin{\pi p\over2}=m_1$
corresponding to $u_b=\pi(1-p)=u_1$ and it is present in
the chanel $e=-$ only.
 
Now we return to the deformed principal model. The S-matrix
of the model, Eq.\ (\ref{Smatrix}), is a tensor product of two copies of
the SG $S$-matrix discussed above with $p=\infty$ in the first factor.
Using the $(Q_L,Q_R)$ labeling of particles, (\ref{Smatrix})
actually means
\bee
S^{(\beta_L^\prime,\beta_R^\prime)
(\alpha_L^\prime,\alpha_R^\prime)}
_{(\alpha_L,\alpha_R)(\beta_L,\beta_R)}(\theta)=
S^{(\infty)\beta_L^\prime \alpha_L^\prime}
_{\alpha_L \beta_L}(\theta)\,
S^{(p)\beta_R^\prime \alpha_R^\prime}
_{\alpha_R \beta_R}(\theta)\,.
\label{Stensor}
\ee
Using a tensor product labeling $E=(e_L;e_R)$ also for the
scattering chanels, the projectors can also be written in the
product form
\bee
\Phi^E_{(\alpha_L,\alpha_R)(\beta_L,\beta_R)}=
\varphi^{e_L}_{\alpha_L \beta_L}\,
\varphi^{e_R}_{\alpha_R \beta_R}\,,
\ee
and similarly for $\Phi_E$. Finally
\bee
S_{(e_L;e_R)}(\theta)=S^{(\infty)}_{e_L}(\theta)
S^{(p)}_{e_R}(\theta)\,.
\ee
 
We are interested in the bound state $B=(0;-)$, which is
composed of the elementary particles (1,1) and (1,-1)
and has lowest mass $m_1$ in this charge (2,0) chanel.
Applying the bootstrap fusion formula (\ref{fusion})
to calculate the scattering of B and one of its
constituents $(1,\gamma_R)$ gives
\bee
S_{B(1,\gamma_R)}(\theta)=
e^{i\delta_\infty(\theta+{iu_1\over2})}
e^{i\delta_\infty(\theta-{iu_1\over2})}S_1^{(p)}(\theta)\,,
\label{SBgamma}
\ee
where
\bee
S_1^{(p)}(\theta)={\sinh\theta+i\sin{\pi(1+p)\over2}\over
\sinh\theta-i\sin{\pi(1+p)\over2}}\,,
\label{S1}
\ee
is the analogous soliton-breather scattering in the SG model.
We can interpret (\ref{SBgamma}) as the product of this SG scattering
coming from the second factor in (\ref{Smatrix}) multiplied by the
diagonal soliton-soliton scattering coming from the first one.
 
Having obtained the bound state-particle scattering, we now
apply (\ref{fusion}) once more to find the $B-B$ scattering
phase, $\delta^{(2,0)}(\theta)$. Using (\ref{SBgamma}) in (\ref{fusion}) we find
\bee
S_{BB}(\theta)=e^{i\delta^{(2,0)}(\theta)}=
e^{2i\delta_\infty(\theta)}
e^{i\delta_\infty(\theta+iu_1)}
e^{i\delta_\infty(\theta-iu_1)}S_{11}^{(p)}(\theta)\,,
\label{SBB}
\ee
where
\bee
S_{11}^{(p)}(\theta)={\sinh\theta+i\sin p\pi\over
\sinh\theta-i\sin p\pi}\,,\qquad u_1=\pi(1-p)=\pi\vert x\vert\,,
\label{S11}
\ee
is the breather-breather scattering phase in the SG model.
The interpretation of (\ref{SBB}) is completely analogous
to that of (\ref{SBgamma}). Finally for later convenience we also give here
the
Fourier representation of $S_{11}^{(p)}(\theta)$:
\bee\label{SGbreather-s}
 S_{11}^{(p)}(\theta)=-\exp\left\{-i\int\limits_{-\infty}^\infty
 d\omega{\sin\omega\theta\over\omega}
 {\cosh(\h-p)\pi\omega\over\cosh\h\pi\omega}\right\}\,.
\ee

\renewcommand{\thesection}{\Alph{section}}
\setcounter{section}{3} \setcounter{equation}{0}
\abschnitt{Appendix C. Perturbative calculations}
 
We chose the following parametrization for the group valued field $G$:
\bee
G=G_0\,{1\over\sqrt{1+\vert \Psi\vert^2}}\pmatrix{1&-\Psi^*\cr \Psi&1\cr}
\pmatrix{e^{-i\Phi}&0\cr0&e^{i\Phi}\cr}\,.
\label{para}
\ee
Here $\Psi$ is a complex scalar field, $\Phi$ is a real scalar field and
$G_0\in SU(2)$ is an appropriate constant group element given by
(\ref{G0BOS}) or (\ref{G0FER}).
 
Before we can expand
the Lagrangian in powers of the coupling constant $\lambda_0$ we have
to rescale our fields as follows.
\bee
\Psi=\sqrt{{\lambda_0\over2}}\,\psi\,,\qquad\qquad
\Phi={1\over2}\sqrt{{\lambda_0\over1+g_0}}\,\phi\,.
\ee
In terms of the rescaled fields we have
\bee
{\cal L}_0={\partial_\mu\psi\partial_\mu\psi^*\over
\big(1+{\lambda_0\over2}\vert\psi\vert^2\big)^2}+{1\over2}
\partial_\mu \phi\partial_\mu \phi -{i\over2}\sqrt{\lambda_0(1+g_0)}
\partial_\mu \phi{\cal A}_\mu-{(1+g_0)\lambda_0\over8}
{\cal A}_\mu{\cal A}_\mu\,,
\label{Lag0ex}
\ee
where
\bee
{\cal A}_\mu=
{\psi\partial_\mu\psi^*-\psi^*\partial_\mu\psi\over
1+{\lambda_0\over2}\vert\psi\vert^2}\,.
\ee
 
Due to the fact that the two cases BOS and FER 
correspond to expansions around different points,
${\cal L}_1$ and ${\cal L}_2$ are different in terms of the
rescaled fields. For the BOS case we find
\bea
{\cal L}_1&=&i\sqrt{2(1+g_0)}h_L\,{\psi+\psi^*\over
1+{\lambda_0\over2}\vert\psi\vert^2}\,
\partial_2\phi+h_L\sqrt{{\lambda_0\over2}}
\,{\vert\psi\vert^2\over
1+{\lambda_0\over2}\vert\psi\vert^2}\,
\big(\partial_2\psi-\partial_2\psi^*\big)\nonumber\\
&+&g_0h_L\sqrt{{\lambda_0\over2}}\,{\psi+\psi^*\over
1+{\lambda_0\over2}\vert\psi\vert^2}\,{\cal A}_2\,,\nonumber\\
\label{Lag1BOS}
\eea
and
\bee
{\cal L}_2=-{2h_L^2\over\lambda_0}-
g_0h_L^2{\big(\psi+\psi^*\big)^2\over
\Big(1+{\lambda_0\over2}\vert\psi\vert^2\Big)^2}\,,
\label{Lag2BOS}
\ee
whereas in the FER case
\bee
{\cal L}_1=2ih_L\sqrt{\lambda_0(1+g_0)}\,\partial_2\phi
\,{\vert\psi\vert^2\over
1+{\lambda_0\over2}\vert\psi\vert^2}
+\Bigg[h_L(1-g_0)-h_R(1+g_0)+
{g_0\lambda_0 h_L\vert\psi\vert^2\over
1+{\lambda_0\over2}\vert\psi\vert^2}\Bigg]{\cal A}_2\,,
\label{Lag1FER}
\ee
and
\bee
{\cal L}_2=-{2(h_L+h_R)^2(1+g_0)\over\lambda_0}+4h_L^2g_0
\,{\vert\psi\vert^2\over
\big(1+{\lambda_0\over2}\vert\psi\vert^2\big)^2}
+4h_Lh_R(1+g_0)
\,{\vert\psi\vert^2\over
1+{\lambda_0\over2}\vert\psi\vert^2}\,.
\label{Lag2FER}
\ee
 
The leading, ${\cal O}(\lambda_0^{-1})$, terms of the free energy are
(\ref{FclassBOS}) and (\ref{FclassFER}).
The next terms are of the order ${\cal O}(1)$ and are given by
half the
logarithmic determinant of the operator corresponding to the \lq free'
Lagrangian, that is the piece quadratic in the fields $\psi$ and $\phi$.
This \lq free' Lagrangian can be written in the BOS case as
\bee
{\cal L}^{(0)}=\partial_\mu\psi^*\,\partial_\mu\psi+
{1\over2}\partial_\mu \phi\,\partial_\mu \phi-g_0h_L^2(\psi+\psi^*)^2+
i\sqrt{2(1+g_0)}h_L(\psi+\psi^*)\,\partial_2\phi\,,
\label{LagfreeBOS}
\ee
and for the FER case as:
\bee
{\cal L}^{(0)}=\partial_\mu\psi^*\,\partial_\mu\psi+
{1\over2}\partial_\mu \phi\,\partial_\mu \phi+\Delta\vert\psi\vert^2+
\omega\,(\psi\partial_2\psi^*
-\psi^*\partial_2\psi)\,.
\label{LagfreeFER}
\ee
 
In both cases, only two fields are really coupled to the external fields.
These are the real part of $\psi$ and $\phi$ in the BOS case and the real
and imaginary parts of $\psi$ in the FER case. This allows us to write
${\cal L}^{(0)}$ as an effective Lagrangian of two bosons $\phi_1$,
$\phi_2$ in both cases:
\bee
{\cal L}^{(0)}_{{\rm eff}}={1\over2}\Big(
\partial_\mu\phi_1\,\partial_\mu\phi_1+\partial_\mu\phi_2\,
\partial_\mu\phi_2\Big)+{m_1^2\over2}\,\phi_1^2+
{m_2^2\over2}\,\phi_2^2+i\nu\,(\phi_2\partial_2\,\phi_1
-\phi_1\partial_2\,\phi_2)\,,
\label{Lageff}
\ee
where
\bee
m_1^2=0\,,\qquad\qquad m_2^2=-4g_0h_L^2\,,\qquad\qquad \nu=\sqrt{1+g_0}\,h_L\,,
\label{paraBOS}
\ee
and
\bee
m_1^2=m_2^2=\Delta\,,\qquad\qquad \nu=\omega\,,
\label{paraFER}
\ee
for the BOS and FER cases, respectively. The one-loop contribution is
half the logarithmic determinant of the linear operator corresponding to
${\cal L}^{(0)}_{{\rm eff}}$:
\bee
f^{(0)}={1\over2}\int{d^np\over(2\pi)^n}{\rm ln}\,
\Big[(p^2+m_1^2)\,(p^2+m_2^2)+4\nu^2p_2^2\Big]\,.
\label{log}
\ee
Taking the derivative of (\ref{log}) with respect to the external fields
and then integrating it again from zero, combined with dimensional analysis
in $n$ dimensions finally gives
\bee
{\cal F}^{(0)}=\delta f^{(0)}={1\over n}\int{d^np\over(2\pi)^n}\,
{(m_1^2+m_2^2)\,p^2+2m_1^2m_2^2+4\nu^2p_2^2\over
(p^2+m_1^2)\,(p^2+m_2^2)+4\nu^2p_2^2}\,.
\label{F0}
\ee
 
We have seen in Section 3 that the 1-loop result in the FER case,
(\ref{F0FER1}), is a function of the sum $h_L+h_R$ only, similarly to
the classical term (\ref{FclassFER}). The free energy calculated by the
TBA method also has this property. It is a very non-trivial fact that
this behaviour can be reproduced by the perturbative calculation.
In fact, we
will now show that in the diagonal case the free energy depends only on the
sum $h_L+h_R$ to all orders of perturbation theory.
 
To show this, we introduce the shifted derivatives
\bee
\tilde\partial_2\psi=\partial_2\psi+\omega\psi\,,
\qquad\qquad{\rm and}\qquad\qquad
\tilde\partial_2\psi^*=\partial_2\psi^*-\omega\psi^*\,.
\label{tilde}
\ee
Using this notation, the
gauged Lagrangian can be compactly written as
\bea
{\cal L}&=&\tilde{\cal L}_0-{2(h_L+h_R)^2(1+g_0)\over\lambda_0}+
iM\sqrt{\lambda_0(1+g_0)}\,
{\vert\psi\vert^2\,\partial_2\phi\,\over
1+{\lambda_0\over2}\vert\psi\vert^2}
\label{Lagtilde}\\
&&+\;{\lambda_0 g_0M\over2}\,
{\vert\psi\vert^2\,\big(\psi\tilde\partial_2\psi^*
-\psi^*\tilde\partial_2\psi\big)\over
\Big(1+{\lambda_0\over2}\vert\psi\vert^2\Big)^2}+
{M^2\vert\psi\vert^2\over
\Big(1+{\lambda_0\over2}\vert\psi\vert^2\Big)^2}\,
\Bigg\{1+{\lambda_0(1-g_0)\vert\psi\vert^2\over2}\Bigg\}\,,
\nonumber\\
\nonumber
\eea
where $\tilde{\cal L}_0$ stands for the original Lagrangian, (\ref{Lag0ex}),
 with all
$\partial_2\psi$ and $\partial_2\psi^*$ substituted by the
shifted derivatives (\ref{tilde}). The advantage of using the form
of the Lagrangian, (\ref{Lagtilde}), is that the explicit dependence on the external
fields $h_L$, $h_R$ occurs only through the combination
$M=(1+g_0)(h_L+h_R)$.
In addition to this, there is still an implicit dependence on the
external fields through $\omega$ hidden in the shifted derivatives
(\ref{tilde}). This disappears, however, since the Lagrangian
(\ref{Lagtilde}) is in fact equivalent to another one, denoted by ${\cal L}_M$,
differing from ${\cal L}$ only by replacing all shifted derivatives,
$\tilde\partial_2$, by ordinary ones. 
 
The fact that ${\cal L}$ and ${\cal L}_M$ are completely equivalent can be
best illustrated in the case of the ${\cal O}(1)$ contribution
to the free energy. The relevant piece of the Lagrangian is
(\ref{LagfreeFER}), which can also be written as
\bee
{\cal L}^{(0)}=\tilde\partial_\mu\psi^*\,
\tilde\partial_\mu\psi+
{1\over2}\partial_\mu \phi\,\partial_\mu \phi+M^2\,\vert\psi\vert^2\,.
\label{LagfreeFER1}
\ee
The substitution $\tilde\partial_2\rightarrow\partial_2$ corresponds to
the shift $p_2\rightarrow p_2+i\omega$ in momentum space. More
precisely, when integrating over $p_2$ in Eq.\ (\ref{F0FER})
the above shift corresponds to deforming the integration contour parallel
to the real axis. This deformation is allowed since the singularities of the
integrand at
\bee
p_2=i\omega\pm i\sqrt{p_1^2+M^2}\,,
\label{sing}
\ee
lie outside the region between the real axis and the shifted contour
(guaranteed by the inegality (\ref{ineq2})).
After shifting the integration variable Eq.\ (\ref{F0FER}) takes the
simple form
\bee
{\cal F}^{(0)}={2\over n}\int{d^np\over(2\pi)^n}\,
{M^2\over
p^2+M^2}\,,
\label{F0FER2}
\ee
which is easily evaluated and yields (\ref{F0FER1}).
Now Eq.\ (\ref{F0FER2}) can be derived from the simplified Lagrangian
\bee
{\cal L}^{(0)}_M=\partial_\mu\psi^*\,
\partial_\mu\psi+
{1\over2}\partial_\mu \phi\,\partial_\mu \phi+M^2\,\vert\psi\vert^2\,,
\label{LagfreeFER2}
\ee
exhibiting the one-loop equivalence between ${\cal L}$ and
${\cal L}_M$. 

It is easy to generalize this argument to show that there is a one-to-one
correspondence between the Feynman diagrams generated by ${\cal L}$ and
${\cal L}_M$ and that the perturbative results are identical for these
two models, diagram by diagram, to all orders of perturbation theory. This
equivalence follows from the fact that the shift
$p_2\rightarrow p_2+i\omega$ can be performed {\sl simultaneously} in all
momentum variables corresponding to the lines of a multi-loop
Feynman diagram. In this way we have shown that all dependence on the
external fields comes from the $M$-dependence of ${\cal L}_M$. This
means, in particular, that the free energy in the diagonal case
takes the form (\ref{FallordI}, \ref{FallordII}) and it is
a function of the sum $h=h_L+h_R$ only, which is exactly the same
as the free energy for the $U(1)_{\rm R}$ case (as a function of $h_R$).
   
\renewcommand{\thesection}{\Alph{section}}
\setcounter{section}{4} \setcounter{equation}{0}
\abschnitt{Appendix D. Renormalization Group Equations}
 
Due to its SU(2)$_{\rm L}\times$U(1)$_{\rm R}$ symmetry, the model (\ref{Lag0})
is renormalizable in perturbation theory. (Of course, in Eq.\ (\ref{Lag0})
actually the bare coupling $\lambda_0$ and the bare $g$-parameter $g_0$
occur.) Using dimensional regularization, the relation between the
bare parameters, $(\lambda_0,g_0)$, and the renormalized ones, $(\lambda,g)$,
is written as:
\bee
\lambda_0=\mu^\epsilon\lambda\,Z_\lambda\,,\qquad\qquad{\rm and}
\qquad\qquad 1+g_0=(1+g)\,Z_g\,.
\label{bare}
\ee
In Eq.\ (\ref{bare}) the dimensionful parameter, $\mu$, is introduced as usual,
to carry
the mass dimension of 
$\lambda_0$ (which is dimensionless in 2 dimensions).
The renormalization constants contain only pole terms in
the variable $\epsilon=2-n$:
\bea
Z_\lambda &=&1+\sum_{r=1}^\infty\,{y_r(\lambda,g)\over\epsilon^r}\,,
\label{Z1}\\
Z_g &=&1+\sum_{r=1}^\infty\,{w_r(\lambda,g)\over\epsilon^r}\,.
\label{Z2}\\
\nonumber
\eea
The residues of the poles, $y_r$ and $w_r$, can be calculated in
perturbation
theory:
\bee
y_r(\lambda,g)\,,\,w_r(\lambda,g)={\cal O}(\lambda^r)\,.
\ee
 
It is clear from (\ref{bare}) that if $g=-1$, then also $g_0=-1$.
This is a consequence of the fact that at this value of the
deformation parameter the model (\ref{Lag0}) reduces to the $O(3)$
nonlinear $\sigma$-model (plus a free boson), which is 
obvious from the parametrization used in Section 4.
An other special value of the deformation coupling is $g=0$ implying
$g_0=0$ and $w_r(\lambda,0)=0$ in Eq.\ (\ref{Z2}).
In this case  the deformation actually vanishes
and we have the 
SU(2)$_{\rm L}\times$SU(2)$_{\rm R}$ (or $O(4)$)
symmetric principal $\sigma$-model. 
In a sense the deformed model (\ref{Lag0}) interpolates
between the $O(3)$ and $O(4)$ nonlinear $\sigma$-models.
 
Physical quantities depend on the renormalized couplings $\lambda$,
$g$ and the dimensionful parameter $\mu$ in such a way that the action
of the renormalization group (RG) operator
\bee
{\cal D}=\mu\,{\partial\over\partial\mu}
+\beta_\lambda(\lambda,g)\,{\partial\over\partial\lambda}
+\beta_g(\lambda,g)\,{\partial\over\partial g}\,,
\label{RGop}
\ee
vanishes on them. Here the two $\beta$-functions are obtained from the
residues of the first order poles in (\ref{Z1}) and (\ref{Z2}) as
\bee
\beta_\lambda=\lambda^2\,{\partial y_1\over\partial\lambda}\,,
\qquad\qquad{\rm and}\qquad\qquad
\beta_g=(1+g)\lambda\,{\partial w_1\over\partial\lambda}\,.
\label{betadef}
\ee
 
For later use we note that in the FER
case the free energy, Eq.\ (\ref{FallordII}), can be written as
\bee
{\cal F}(h)=-2h^2\,\Bigg\{{1+g_0\over\lambda_0}-{(1+g_0)^2\over2\lambda_0}\,
S(g_0,\lambda_0 \big[h(1+g_0)\big]^{-\epsilon})\Bigg\}\,,
\label{Fallord1}
\ee
where $S$ stands for the infinite sum in (\ref{FallordII}). From 
(\ref{Fallord1}) we can conclude that at $g=-1$ $Z_\lambda=Z_g$
and in particular,
\bee
w_1(\lambda,-1)=y_1(\lambda,-1)\,.
\label{wy}
\ee
 
In both BOS and FER cases, the free energy takes the form
\bee
{\cal F}(h)=-h^2\,F_0(\lambda,g,\mu,h)\,,
\label{free}
\ee
where the function $F_0$ is RG invariant: ${\cal D}F_0=0$.
As we are interested in the asymptotic expansion of the free energy for
large external fields we write
\bee
h=h_0\,e^t\,,
\label{h0}
\ee
where $h_0$ is a fixed, finite value of the external field and
$t\rightarrow\infty$.
Standard RG considerations then show that
\bee
{\cal F}(h)=-h^2\,F_0(\bar\lambda(t),\bar g(t),\mu,h_0)\,,
\label{RG}
\ee
where the running coupling $\bar\lambda$ and the running deformation
parameter $\bar g$ satisfy the following set of differential equations
and initial conditions.
\bea
{d\bar\lambda\over dt} =\beta_\lambda(\bar\lambda,\bar g)\,,&
\qquad\qquad
&\bar\lambda(0)=\lambda\,,\label{RGeq1}\\
{d\bar g\over dt} =\beta_g(\bar\lambda,\bar g)\,,&
\qquad\qquad\,
&\bar g(0)=g\,.\label{RGeq2}\\
\nonumber
\eea
 
In the deformed principal model the $\beta$-functions are known to be of
the form \cite{Mouchanna,beta-fun}
\bea
\beta_\lambda&=& -{\lambda^2\over4\pi}\Bigg\{1-g+\lambda p_2(g)+
\lambda^2 p_3(g)+\cdots\Bigg\}\,,\label{beta1}\\
\beta_g&=& {\lambda g(1+g)\over2\pi}\Bigg\{1+\lambda q_1(g)+
\lambda^2 q_2(g)+\cdots\Bigg\}\,,\label{beta2}\\
\nonumber
\eea
where the two-loop $\beta$-function coefficients are given by 
\cite{Mouchanna, beta-fun} 
\bee
p_2(g)={1-2g+5g^2\over8\pi}\,,\qquad\qquad{\rm and}
\qquad\qquad q_1(g)={1-g\over4\pi}\,.
\label{2loop}
\ee
The three-loop coefficients $p_3(g)$, $q_2(g)$ are not known at present,
but fortunately we will not need their explicit form beyond the combination
\bee
u=2q_2(-1)-p_3(-1)\,.
\label{u}
\ee
  
We are interested in solutions of (\ref{RGeq1}) and (\ref{RGeq2})
with asymptotically free
(AF) behaviour, that is $\bar\lambda
\rightarrow0$ together with $\bar g\rightarrow g_1$ as $t\rightarrow\infty$,
where
$g_1$ is some constant.
Analyzing Eqs.\ (\ref{beta1}), (\ref{beta2}), we find
three different AF solutions:
\bea
{\rm 1.}\qquad &g_1=\phantom{-}0&
\qquad\qquad g\equiv g_0\equiv\bar g\equiv0\,,
\label{AF1}\\
{\rm 2.}\qquad &g_1=-1&\qquad\qquad g\equiv g_0\equiv\bar g\equiv -1\,,
\label{AF2}\\
{\rm 3.}\qquad &g_1=-1&\qquad\qquad -1<\bar g<0\,.
\label{AF3}\\
\nonumber
\eea
Solutions (\ref{AF1}) resp.\ (\ref{AF2}) correspond to the O(4) resp.\ 
the O(3) NLS model, while the solution (\ref{AF3}) corresponds to
the \lq generic' case of deformed principal model. In what follows
we shall concentrate on the generic case. We introduce the RG-invariant
combination of the two couplings:
\bee
p=2\pi\lim_{t\rightarrow\infty} {1+\bar g(t)\over\bar\lambda(t)}\,,
\label{g}
\ee
which we {\sl identify} with the parameter $p$ in the $S$ matrix 
(\ref{Smatrix}).
A RG-invariant quantity can almost be treated just as 
a numerical parameter. We can express the running deformation coupling,
$\bar g$, in terms of the running coupling, $\bar\lambda$
and the RG-invariant $p$ as
\bee
\bar g=\Gamma(\bar\lambda,p)
\label{Gamma}
\ee
and define an effective $\beta$-function for $\lambda$ as
\bee
\beta_{\scriptscriptstyle{\rm eff}}(\lambda,p)
=\beta_\lambda(\lambda,\Gamma(\lambda,p))\,.
\label{betaeff}
\ee
Using the perturbative expressions (\ref{beta1}) and (\ref{beta2})
together with (\ref{2loop}) and (\ref{u}) we have
\bee
\Gamma(\lambda,p)=-1+{p\over2\pi}\lambda +
p^{(2)}\lambda^2+p^{(3)}\lambda^3+\cdots\,,
\label{Gammaexp}
\ee
where
\bee
p^{(2)}=-{p^2\over8\pi^2}\,,\qquad\qquad{\rm and}\qquad\qquad
p^{(3)}={p^2(p+2)\over64\pi^3}+{pu\over8\pi}\,.
\label{p2p3}
\ee
The unknown three-loop coefficients contribute only to $u$
defined in Eq.\ (\ref{u}). It is not difficult to show using Eq.\ (\ref{wy}) 
that
\bee
2q_2(-1)=p_3(-1)\,,
\ee
implying $u=0$.

Using the perturbative form (\ref{Gammaexp}) in the definition
(\ref{betaeff}) we have
\bee
\beta_{\scriptscriptstyle{\rm eff}}(\lambda,p)=-{\lambda^2\over2\pi}+
{p-2\over8\pi^2}\lambda^3+\cdots\,,
\label{betaeff1}
\ee
which allows the definition of the RG-invariant $\Lambda$-parameter
in the $\overline{\rm MS}$ scheme the usual way:
\bee
\Lambda_{\overline{\rm MS}}=\mu\, e^{-{2\pi\over\lambda}}
\Big({\lambda\over
2\pi}\Big)^{\big({p\over2}-1\big)}
\,e^\gamma\Big\{1+{\cal O}(\lambda)\Big\}\,.
\label{Lambda}
\ee
 
The next step in the RG analysis is the introduction of the effective
coupling $\eff(h)$ defined by the transcendental equation
\bee
{2\pi\over \eff}+\Big({p\over2}-1\Big)\ln{2\pi\over \eff}=\ln{h\over
\Lambda_{\overline{\rm MS}}}\,.
\label{eff}
\ee
The advantage of using the effective charge is that it is a function
of the physical quantity
\bee
s=\ln{h\over\Lambda_{\overline{\rm MS}}}\,,
\label{s}
\ee
only and moreover the running coupling can be expressed in terms of
$\eff(h)$ perturbatively (in the sense that it is an infinite power
series):
\bee
\bar\lambda=\eff+{1\over2\pi}\Big(\ln{h_0\over\mu}-
\gamma\Big)\eff^2+\cdots\;.
\label{lambdabar}
\ee
The asymptotic expansion (for large $s$) of the
effective coupling (containing terms $\propto\ln s$) can be written as 
\bee
\eff={2\pi\over s}+{\pi(p-2)\over s^2}\ln s+\cdots\;.
\label{e1}
\ee
 
So far we have concentrated on the generic case (\ref{AF3}). The O(3) 
symmetric
limit, case (\ref{AF2}), ($p\to0$) corresponds to putting simply $p=0$ 
in Eq.\ (\ref{e1}).
The O(4) limit, (\ref{AF1}), however, cannot be obtained from the
generic case although it corresponds formally to the $p\rightarrow\infty$
limit. (See the discussion in Subsection 2.3.)
 
For lattice regularization some of the above formulae are modified.
First of all, instead of (\ref{bare}) and (\ref{Z1}-\ref{Z2}) 
the relation between the
bare parameters $(\tilde{\lambda}_0,\tilde{g}_0)$ and the
renormalized ones $(\tilde{\lambda},\tilde{g})$ are given by
\bee
\tilde{\lambda}_0=\tilde{\lambda}\,\tilde{Z}_\lambda\,,
\qquad\qquad{\rm and}
\qquad\qquad 1+\tilde{g}_0=(1+\tilde{g})\,\tilde{Z}_g\,,
\label{bareLatt}
\ee
where 
\bea
\tilde{Z}_\lambda &=&1+\sum_{r=1}^\infty\,
\tilde{y}_r(\tilde{\lambda},\tilde{g})\,(\ln \tilde{\mu}a)^r\,,
\label{Z1Latt}\\
\tilde{Z}_g &=&1+\sum_{r=1}^\infty\,
\tilde{w}_r(\tilde{\lambda},\tilde{g})\,(\ln \tilde{\mu}a)^r\,.
\label{Z2Latt}\\
\nonumber
\eea
Here the parameter $\tilde{\mu}$ is the lattice analogue of the
$\mu$-variable of the dimensional scheme and naturally it is
$\tilde{\mu}$ that appears in the lattice version of (\ref{RGop}).
The lattice $\beta$-functions are given by
\bee
\tilde{\beta}_\lambda=-\tilde{\lambda}\,
\tilde{y}_1(\tilde{\lambda},\tilde{g})\,,
\qquad\qquad{\rm and}\qquad\qquad
\tilde{\beta}_g=-(1+\tilde{g})\,
\tilde{w}_1(\tilde{\lambda},\tilde{g})\,.
\label{betadefLatt}
\ee

The rest of the RG analysis is identical to the one in the dimensional
scheme. Although the higher order $\beta$-functions are different,
the main features of the RG trajectories (\ref{AF1}-\ref{AF3}) remain the
same and also the first two coefficients of 
$\beta_{{\scriptscriptstyle {\rm eff}}}(\tilde{\lambda},p)$ are unchanged. 
The lattice version of the $\Lambda$-parameter is given by
\bee
\Lambda_{L}=\tilde{\mu}\, e^{-{2\pi\over\tilde{\lambda}}}
\Big({\tilde{\lambda}\over
2\pi}\Big)^{\big({p\over2}-1\big)}
\,\Big\{1+{\cal O}(\tilde{\lambda})\Big\}\,.
\label{LambdaLatt}
\ee

%-----------------------------------------

%{\it Acknowledgements}


\newpage
\begin{thebibliography}{99}
%
\bibitem{PW}
A.~Polyakov, P.B.~Wiegmann,
Phys. Lett. 131B (1983) 121
%
\bibitem{Wiegmann}
P.B.~Wiegmann,
Phys. Lett. 152B (1985) 209.
%
\bibitem{Cherednik}
I.V.~Cherednik
Sov. J. Nucl. Phys., 33 (1981) 144
%
\bibitem{Babu}
H.M.~Babujian, A.M.~Tsvelick,
Nucl. Phys. B265 (1986) 24.
%
\bibitem{Kirillov}
A.~Kirillov, N.Yu.~Reshetikhin in Proc. of the Paris-Meudon
Colloquium,
String Theory, Quantum Cosmology and Quantum Gravity, Integrable
and Conformal Invariant Theories, (1986), eds.\ N. Sanchez, H.~ de Vega,
(World Scientific, Singapore).
%
\bibitem{Fateev}
V.A.~Fateev,
Nucl. Phys. B473[FS] (1996) 509.
%
\bibitem{hollo}
J.M.~Evans and T.J.~Hollowood,
Nucl. Phys. Proc. Suppl. 45A (1996) 130; 
%
\bibitem{Jap}
G. Japaridze, A. Nersesyan and P. Wiegmann,
Nucl. Phys. B230 (1984) 511.
%
\bibitem{history}
P.~Hasenfratz, M.~Maggiore and F.~Niedermayer,
Phys. Lett. 245B (1990) 522; ibid 529.
%
\bibitem{FoNiWe}
P.~Forg\'acs, F.~Niedermayer and P.~Weisz,
Nucl. Phys. B367 (1991) 123.
%
\bibitem{fendley}
P.~Fendley and K.~Intriligator,
Phys. Lett. 319B (1993) 132; 
%
\bibitem{Balog}
J.~Balog, S.~Naik, F.~Niedermayer and P.~Weisz,
Phys.~Rev.~Lett. 69 (1992) 873.
%
\bibitem{Karowski}
M.~Karowski,
Nucl.~Phys.~B153 (1979) 244.
% 
\bibitem{Mouchanna}
P. Azaria, P. Lecheminant and D. Mouhanna,
Nucl. Phys. B455 (1995) 648;
%
\bibitem{beta-fun}
J.~Balog, P.~Forg\'acs, Z.~Horv\'ath and L.~Palla,
Nucl. Phys. Proc. Suppl. 49B (1996) 16;
%
\bibitem{Karp}
L. Karp and L.~Palla,
to be published
%
\bibitem{Zamol}
A.B.~Zamolodchikov and A.B.~Zamolodchikov,
Ann.\ Phys.\ (NY) 120 (1979) 253
%
\bibitem{Coleman}
S.~Coleman and H.J.~Thun, Comm. Math. Phys. 61 (1978) 31
%
%
\bibitem{Karowski2}
H.~Babujian, A.~Fring, M.~Karowski and A.~Zapletal,
Nucl. Phys. B538 (1999) 535
%
%
\bibitem{Mussardo}
G.~Mussardo, Phys. Rep. 218 (1992) 215
%
 
%---------------------
 
 
\end{thebibliography}
\end{document}